\documentclass[sigconf]{acmart}

\usepackage[utf8]{inputenc} % allow utf-8 input
\usepackage[T1]{fontenc}    % use 8-bit T1 fonts
\usepackage{hyperref}       % hyperlinks
\usepackage{url}            % simple URL typesetting
\usepackage{amsfonts}       % blackboard math symbols
\usepackage{nicefrac}       % compact symbols for 1/2, etc.
\usepackage{microtype}      % microtypography
\usepackage{xcolor}         % colors
\usepackage{siunitx}
\usepackage{graphicx,subcaption}
\usepackage{amsmath}
\usepackage{enumitem}
\usepackage{multirow}
\usepackage{balance}
\usepackage{listings}
\usepackage{acmart-taps}

\aptLtoXcmd{}
{\definecolor{codegreen}{rgb}{0,0.6,0}
\definecolor{codegray}{rgb}{0.5,0.5,0.5}
\definecolor{codepurple}{rgb}{0.58,0,0.82}
\definecolor{backcolour}{rgb}{0.95,0.95,0.92}
\lstdefinestyle{mystyle}{
    backgroundcolor=\color{backcolour},   
    commentstyle=\color{codegreen},
    keywordstyle=\color{magenta},
    numberstyle=\tiny\color{codegray},
    stringstyle=\color{codepurple},
    basicstyle=\ttfamily\footnotesize,
    breakatwhitespace=false,         
    breaklines=true,                 
    captionpos=b,                    
    keepspaces=true,                 
    numbers=none,                    
    numbersep=5pt,                  
    showspaces=false,                
    showstringspaces=false,
    showtabs=false,                  
    tabsize=2
}
\lstset{style=mystyle}}

\newcommand{\xref}[1]{\S\ref{#1}}

\DeclareMathOperator*{\argmax}{arg\,max}

\newenvironment{todo-env}{\par\color{red}}{\par}
\newenvironment{help-env}{\par\color{blue}}{\par}
\newenvironment{ready-for-review}{\par\color{violet}}{\par}

\copyrightyear{2024}
\acmYear{2024}
\setcopyright{rightsretained}
\acmConference[UIST '24]{The 37th Annual ACM Symposium on User Interface Software and Technology}{October 13--16, 2024}{Pittsburgh, PA, USA}
\acmBooktitle{The 37th Annual ACM Symposium on User Interface Software and Technology (UIST '24), October 13--16, 2024, Pittsburgh, PA, USA}
\acmDOI{10.1145/3654777.3676327}
\acmISBN{979-8-4007-0628-8/24/10}

\setlist[itemize]{leftmargin=*}

\begin{document}

\title{{IRIS: Wireless ring for vision-based smart home interaction}}

 \author{Maruchi Kim}\authornote{Corresponding authors}
  \affiliation{Paul G. Allen School, University of Washington
  \city{Seattle, WA}
 \country{USA}
 }
 \email{mkimhj@cs.washington.edu}

 \author{Antonio Glenn}\authornotemark[1]
  \affiliation{Paul G. Allen School, University of Washington
  \city{Seattle, WA}
 \country{USA}
 }
 \email{aglenn5@cs.washington.edu}

 \author{Bandhav Veluri}\authornotemark[1]
  \affiliation{Paul G. Allen School, University of Washington
  \city{Seattle, WA}
 \country{USA}
 }
 \email{bandhav@cs.washington.edu}

 \author{Yunseo Lee}
  \affiliation{Paul G. Allen School, University of Washington
  \city{Seattle, WA}
 \country{USA}
 }
 \email{yunseol1@cs.washington.edu}

 \author{Eyoel Gebre}
  \affiliation{Paul G. Allen School, University of Washington
  \city{Seattle, WA}
 \country{USA}
 }
 \email{eyoel23@cs.washington.edu}

 \author{Aditya Bagaria}
  \affiliation{Paul G. Allen School, University of Washington
  \city{Seattle, WA}
 \country{USA}
 }
 \email{abagaria@cs.washington.edu}

 \author{Shwetak Patel}
  \affiliation{Paul G. Allen School, University of Washington
  \city{Seattle, WA}
 \country{USA}
 }
 \email{shwetak@cs.washington.edu}

 \author{Shyamnath Gollakota}\authornotemark[1]
  \affiliation{Paul G. Allen School, University of Washington
  \city{Seattle, WA}
 \country{USA}
 }
 \email{gshyam@cs.washington.edu}

\renewcommand{\shorttitle}{IRIS: Wireless ring for vision-based smart home interaction}
\renewcommand{\shortauthors}{Kim et al.}

%\pagenumbering{arabic}

\begin{abstract}

Integrating cameras into wireless smart rings has been challenging due to size and power constraints. We introduce IRIS, the first wireless vision-enabled smart ring system for smart home interactions. Equipped with a camera, Bluetooth radio, inertial measurement unit (IMU), and an onboard battery, IRIS meets the small size, weight, and power (SWaP) requirements for ring devices. IRIS is context-aware, adapting its gesture set to the detected device, and can last for 16-24 hours on a single charge. IRIS leverages the scene semantics to achieve instance-level device recognition. In a study involving 23 participants, IRIS consistently outpaced voice commands, with a higher proportion of participants expressing a preference for IRIS over voice commands regarding toggling a device's state, granular control, and social acceptability.  Our work pushes the  boundary of what is possible with ring form-factor devices, addressing system challenges and opening up novel interaction capabilities.

\end{abstract}

\begin{CCSXML}
<ccs2012>
   <concept>
       <concept_id>10003120.10003121</concept_id>
       <concept_desc>Human-centered computing~Human computer interaction (HCI)</concept_desc>
       <concept_significance>500</concept_significance>
       </concept>
   <concept>
       <concept_id>10003120.10003138</concept_id>
       <concept_desc>Human-centered computing~Ubiquitous and mobile computing</concept_desc>
       <concept_significance>500</concept_significance>
       </concept>
   <concept>
       <concept_id>10010147.10010257</concept_id>
       <concept_desc>Computing methodologies~Machine learning</concept_desc>
       <concept_significance>500</concept_significance>
       </concept>
   <concept>
       <concept_id>10010520.10010553.10010562</concept_id>
       <concept_desc>Computer systems organization~Embedded systems</concept_desc>
       <concept_significance>500</concept_significance>
       </concept>
    <concept>
       <concept_id>10010583.10010786</concept_id>
       <concept_desc>Hardware~Emerging technologies</concept_desc>
       <concept_significance>500</concept_significance>
       </concept>
 </ccs2012>
\end{CCSXML}

\ccsdesc[500]{Human-centered computing~Human computer interaction (HCI)}
\ccsdesc[500]{Human-centered computing~Ubiquitous and mobile computing}
\ccsdesc[500]{Computing methodologies~Machine learning}
\ccsdesc[500]{Computer systems organization~Embedded systems}
\ccsdesc[500]{Hardware~Emerging technologies}

\keywords{Smart ring, context-aware interaction, low-power cameras, efficient deep learning, smart homes, IoT, wearables}

\begin{teaserfigure}
    \centering
    \includegraphics[width=\textwidth]{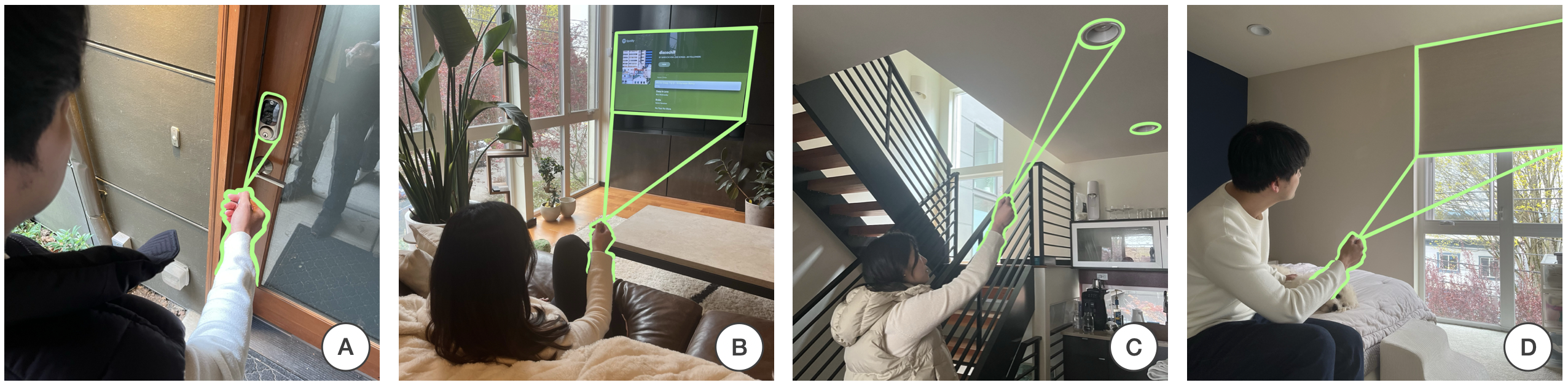}
    \caption{Smart home interaction with IRIS. (A) A user unlocks the front door by pointing and clicking IRIS at the smart lock. (B) Another user points IRIS at a television and rotates their hand to adjust its volume. (C) The user points IRIS at their living room lights to turn them off before leaving home. (D) A user points and clicks IRIS at the blinds to lower them for privacy.}
    \label{fig:banner}
    \Description{4 photos labeled (A), (B), (C), and (D) show people using IRIS to interact with various smart home devices. Image (A) shows the back of a man pointing IRIS at a smart lock to unlock his front door. Image (B) shows the back of a woman in her living room pointing IRIS at a television to control it. Image (C) shows the same woman wearing a vest, pointing at her ceiling lights to turn them on. Image (D) shows the same man from image (A) in his bedroom, pointing IRIS at a set of blinds to raise lower them.}
\end{teaserfigure}

\maketitle

\section{Introduction}

As households transition into smart homes, they are outfitted with an array of interconnected devices, like smart speakers, door locks, and other smart home appliances~\cite{retaildive_2024, oberlo_2024}. While these devices promise improved convenience, the means by which users control them remain ripe for improvement.  Issues such as social discomfort in using voice commands and the unreliability of voice input in noisy environments hinder seamless interaction~\cite{JAIN2023102662, 10.1145/3335595.3335635}. Despite the availability of smartphone apps for direct control, it is often more convenient to resort to traditional methods, such as light switches or dedicated remotes, rather than unlocking one's phones, locating the appropriate application, and navigating through its interface \cite{fastcompany2022, brillianttech2022}. 

%In this work, we pose the following question:Can a our smart rings could ``see'' what we want to control, overcoming the limitations of auditory commands? 

%This paper aims to tap into the under-utilization of object recognition in wearables by proposing a novel system for controlling smart home devices -- a wireless, camera-enabled ring for smart home interaction. 

We introduce IRIS, short for Interactive Ring for Interfacing with Smart home devices. IRIS is an end-to-end wireless ring system  that supports real-time object instance detection using contextual scene semantics. The underlying principle is rooted in the age-old adage that a ``picture is worth a thousand words,'' asserting that capturing images is far more efficient than verbalizing lengthy auditory commands. IRIS enables users to control smart home devices by simply pointing at the target device and performing a corresponding gesture, offering an intuitive alternative to traditional interaction methods. {Finally, IRIS requires no additional setup beyond standard smart home device installation, and the experiments and results presented in this work were all conducted on existing smart home devices with no  modification.}

%designed to offer a more efficient and socially acceptable interaction modality, while leveraging the potential of real-time object detection. IRIS introduces a novel approach to smart home device interaction -- a wireless, camera-enabled ring driven by real-time instance detection and a human-in-the-loop learning system to surpass voice commands in usability and efficiency. The underlying principle is rooted in the age-old adage that a "picture is worth a thousand words," asserting that sending images is far more efficient than verbalizing lengthy auditory commands. This key insight led to the creation of IRIS, enabling users to control smart home devices by simply pointing at the target device and inputting a corresponding gesture, offering a swift and intuitive alternative to traditional interaction methods.

%Thus, this paper introduces a novel low-power design to enable efficient video streaming in a smart ring form factor.

Achieving this is challenging for three key reasons. First, while sensors integrated in today's ring devices, such as IMUs, are low-power, camera hardware can generate significantly more data, leading to orders of magnitude higher power consumption~\cite{neuricam}. It is unclear if wireless camera systems can be designed to meet the small size, weight and power requirements (SWaP) of ring form-factor devices. Secondly, while object detection systems excel in real-time detection, their limitations become evident when confronted with multiple instances of the same object class. For instance, merely determining that a user pointed at a set of blinds is inadequate; the system must also discern which specific set of blinds in the home the user intends to control. Therefore, a system capable of precisely identifying and distinguishing between individual devices of the same class is necessary. Finally, the entire end-to-end system should operate in real-time under around one second  to be considered a seamless interaction modality ~\cite{NielsenResponseTimes, endo1996using}.

% Finally, given the diverse nature of homes and the continuous influx of new smart home devices, the system must generalize across various environments and lighting scenarios while also adapting to emerging Internet-of-Things (IoT) devices. 

In IRIS, we address these challenges by making technical contributions spanning hardware, software, and system design. At a high level, the wearer points at the device of interest and presses the button on the ring, as shown in Fig.~\ref{fig:banner}. This wakes up the on-board camera which captures a few frames of the target device, which are  streamed to a nearby smartphone for processing via Bluetooth. The wearer also performs a gesture (e.g. rotation) to control the device, the intent of which is captured using the on-board IMU. This is also transmitted to the phone which then, in real-time, controls the target device. Specifically, we make three key contributions. 

\begin{figure}[t!]
\vskip -0.1in
\centering
\includegraphics[width=1\linewidth]{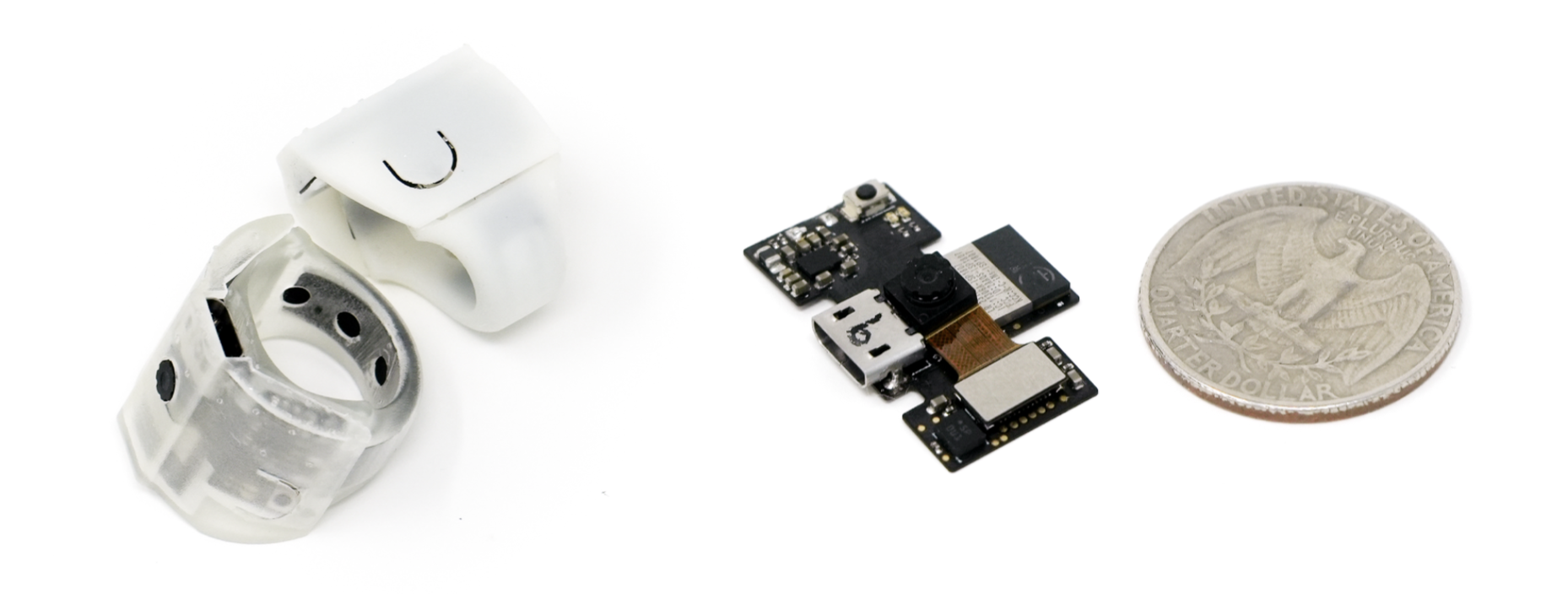}
\vskip -0.1in
\caption{\textmd{IRIS hardware inside 3D-printed enclosure and when placed beside a quarter. The battery sits inside the band of the ring. The ring diameter and band thickness are 17.5 and 2.9 mm.}}
\Description{This image contains two IRIS rings, IRIS printed circuit board, and a US quarter. There is a clear enclosure of IRIS showing the camera on the left along with a USB connector. The matte white IRIS ring a profile to view the button. To the right of the image is a US quarter, placed beside the assembled IRIS circuit board.}
\label{fig:iris}
\vspace*{-10pt}
\end{figure}

%\squishlist
\begin{itemize}
\item {\bf SWAP-constrained wireless camera ring.} We designed the first wireless ring
form-factor device for vision-based smart home interaction (Fig.~\ref{fig:iris}). Our hardware is equipped with a camera, Bluetooth radio, IMU, and an on-board battery, while meeting the small size, weight and power (SWaP) requirements expected in ring form-factor devices. {We present cross-layer optimization methods across wireless camera hardware and firmware, and a user interface that enables extremely low-power states.} We designed our DIY hardware using open-source eCAD software, outsourced fabrication and assembly, and 3D printed the enclosures. The final fabrication cost (PCB and components) was \$471 for 20 units. The PCB, battery, and enclosure weigh 4 grams. {IRIS is the first ring to stream camera data wirelessly, and operate for over 16 hours on a single charge.}% in total. 

%The final bill of materials (BOM for the components and PCB were \$28.07.

\item {\bf Instance-level classification based on scene semantics.} 
{Our machine learning
 (ML) pipeline surpasses traditional object detection models by enabling scene-based understanding and detection of specific instances within a class, all on resource-constrained devices.} To achieve instance-level classification, we start by utilizing a self-supervised vision transformer model, DINOV2 \cite{oquab2024dinov2}, to generate scene-level embeddings. These embeddings capture not only the object itself but also the surrounding environment, providing valuable context and scene semantics. However, DINOV2's search runtime increases linearly with the size of the embedding database, introducing potential latency issues for an interactive system like IRIS. To address this challenge, we reduce the search space of DINOV2 by utilizing YOLO \cite{redmon_2016} to first detect all potential smart home devices within the image frame. Since YOLO can detect multiple objects, we use a centered-object detection algorithm (CODA). This analyzes the bounding boxes generated by YOLO and selects the object (smart device) closest to the image center. The classification from CODA serves to significantly reduce the search space for DINOV2's query. Our results demonstrate that this optimization effectively reduces DINOV2's query runtime by hundreds of milliseconds.

% \textcolor{red}{we start off with Dinov2 for xxx, however the dino runtime goes up with the number of references. To alleviate this, we use YOLO to first classify the object, and use CODA to output a single class to limit the search spaceof dino blah blah say dino is a self-supervised vision transformer model}. We use both YOLO~\cite{xx} and DINOV2~\cite{xx}  to address the multitude of scenarios that may arise when using a vision-based approach to control a smart home. Since our wireless ring system may detect multiple smart home devices within a single frame,  we deploy a centered object detection algorithm (CODA~\cite{xxx}), which utilizes the bounding boxes from YOLO to output a single class corresponding to the object closest to the center of the image. In homes with multiple instances of the same class, we utilize DINOV2 to understand not only the detected object, but the entire scene in which it is located. We use these scene semantics to identify specific smart device instance in the user's home.

% Voice assistants continuously learn and improve based on the characteristics of the user's speech \cite{sun_2019}. IRIS employs a human-in-the-loop learning-based system that learns with usage as well. Our system learns to correct errors through a user-initiated feedback mechanism.

\item {\bf End-to-end system optimization for real-time operation and low-power.} Real-time operation is critical for interactive mobile systems \cite{NielsenResponseTimes, endo1996using}. We optimize our wireless ring's streaming performance to maximize camera throughput, and show that the end-to-end system can control devices within one second, delivering near real-time feedback to users by confirming successful gesture recognition. Furthermore, we extensively optimized IRIS's low-power design for all day use, despite the limited on-board battery capacity.

% To tackle the issue of generalizability and determining which particular device a user wants to control, we introduce an interactive learning-based system. We start with a DinoV2 foundation model to capture image-level embeddings and patch-level embeddings. \textcolor{red}{mention coda} \textcolor{red}{Bandhav to add more detail here}.

%\squishend
\end{itemize}
% \textcolor{red}{(Do we need this next part? In failure scenarios, users can simply double-press the IRIS's button to (1) Undo the most recent action taken by our system, and (2) Create a new reference image so that the vision system learns from its error and doesn't repeat the same mistake twice.)} 

Put together, we design an end-to-end wearable ring system capable of (1) smart home device interaction that complements voice commands, (2) instance-based object detection to correctly distinguish between individual instances of the same class, and (3) real-time operation, on-device processing on a standard mobile phone.  We evaluate our system in five different homes from various angles and lightning conditions. Our real-world dataset also includes 98 examples of blinds, 57 doors, 34 door handles, 228 lights, 24 smart locks, 73 speakers, 64 televisions, 37 windows, and 161 background instances. Our results show that:

\begin{itemize}
%\squishlist
\item Our wireless ring hardware can stream video at 3.4 frames per second to a phone, supports context-aware gestures, and can operate for 16-24 hours on a rechargeable 27mAh battery.
\item A user study with 23 participants who performed 690 interactions with our ring hardware shows that participants generally favored IRIS over a voice assistant for controlling devices and greater social acceptability. IRIS performs in real-time, outpacing voice commands by 2 seconds on average, from command initiation.

\item  Our system enables instance-level detection, while optimizing for runtime performance. Across 18 unique instances of devices, including 2 HVACs, 2 blinds, 1 door, 4 lighting systems, 2 smart locks, 5 speakers, and 2 TVs, the instance-based classification accuracy was 95\% and 98\% when provided with 2 and 3 reference images, respectively. Further, we show an average inference latency reduction from 411ms for the DINOV2 model to 112ms for our DINO+YOLO+CODA system.

%\squishend
\end{itemize}

We believe that this paper introduces a novel approach to smart home device interaction through IRIS -- a wireless, low-power, camera-enabled ring backed by real-time instance detection and contextual scene semantics. By working across both hardware and software, IRIS provides an alternate modality to existing voice command and app-based interactions. Our results highlight IRIS's viability, showcasing its real-time responsiveness and user experience {achieved with optimization techniques across hardware, firmware, and ML runtime}. With the potential to impact the landscape of ring-based human-computer interaction, we believe that IRIS has the potential to be an important step toward a more natural, unobtrusive, and user-friendly interface for smart homes.

% Our results highlight IRIS's viability, showcasing its real-time responsiveness and user experience, and adaptability to both new and unseen IoT devices. With the potential to redefine the landscape of ring-based human-computer interaction, we believe that IRIS represents an important step toward a more natural, unobtrusive, and user-friendly interface for smart homes.

\begin{figure}[t!]
\centering
    \includegraphics[width=0.49\textwidth]{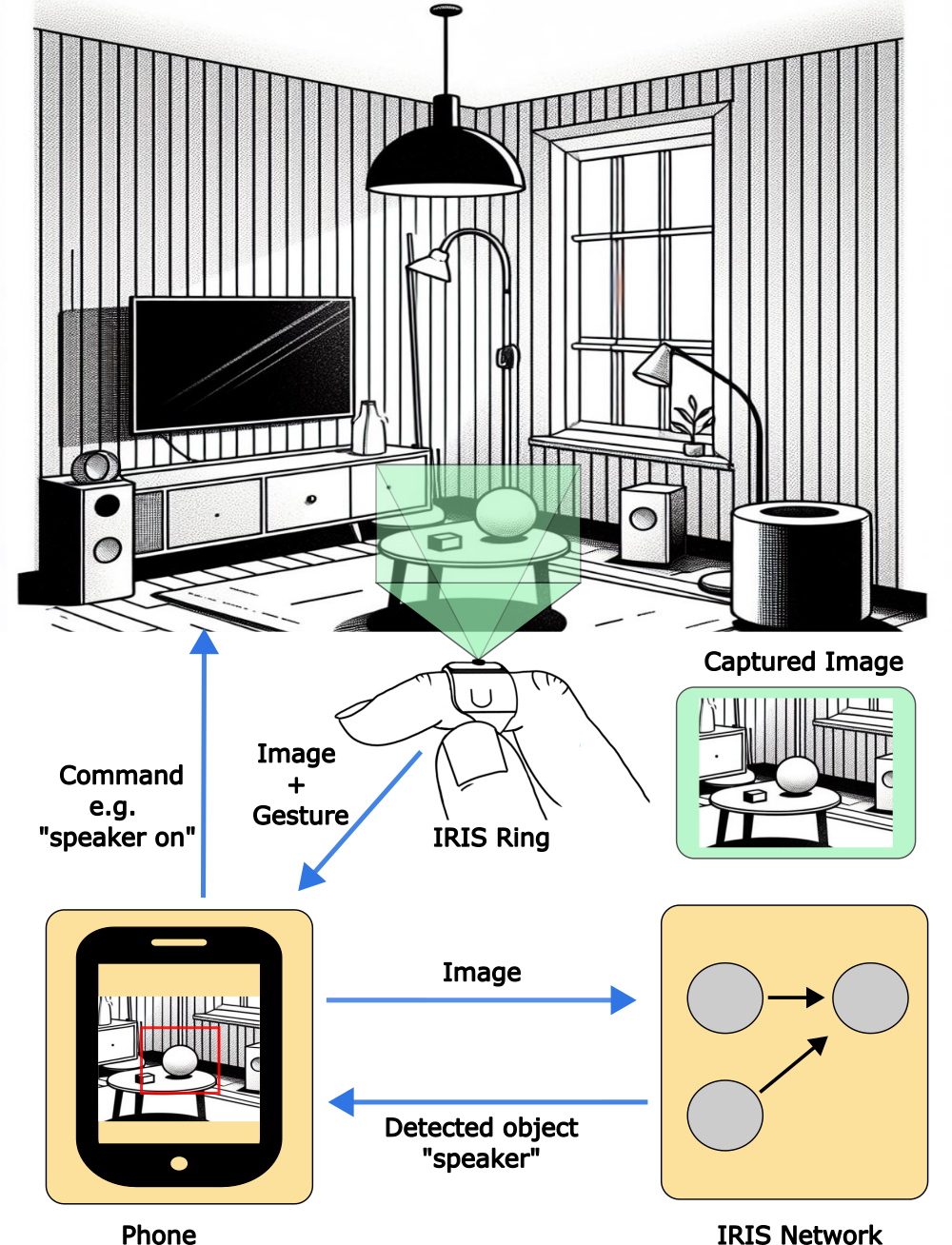}
\caption{{\bf IRIS in Action: Context-aware smart home control. \textmd{This figure illustrates the application of IRIS, a custom-designed, camera-enabled wireless ring for context-aware control of the smart home. IRIS utilizes instance-based object detection for real-time interaction with the environment. }}        }
\Description{Graphical diagram of the IRIS system. In the center of the image, IRIS ring is shown on a user’s finger pointing at an illustration of a smart speaker in a living room. The captured image is illustrated in a green box to the right of the diagram. This image, shown as an illustration, is shown being transmitted to a phone. The image is sent to the yellow block on the right which is our neural network. The neural network (bottom right yellow box) returns the class of the object, a speaker here, and the phone issues commands to turn the speaker on.}
\label{fig:system-overview}
\vskip -0.15in
\end{figure}

\section{Related Work}

To the best of our knowledge, we present the first wireless ring form-factor system for vision-based smart home interaction. Below we present related works across IoT and ring prototypes.

\noindent{\bf Voice interaction for IoT devices.}  A common modality for interaction with Internet-of-Things (IoT) devices in smart-homes continues to be through voice-based IoT devices such as Google Home \cite{HowSetSmart}, Apple Home Pod \cite{HomePod}, and Amazon Alexa \cite{AmazonAlexaVoice}. As noted in \cite{MinuetMultimodalInteractionKangEtAl2019a}, as the number of IoT devices in the home continues to grow, using  voice, and associated apps, to control these object places a burden on the mental load of users which limits their scale.

%We build on this work but instead focus on designing the first wireless ring for vision-based home interaction and address the small weight and power (SWaP) requirements of such wireless devices.

\noindent{\bf Cameras for IoT interaction.}  Cameras can enable contextual understanding of gestures and intent.  Snap-to-It \cite{SnapToItUserInspiredPlatformdeFreitasEtAl2016} and SnapLink \cite{SnapLinkFastAccurateChenEtAl2018},  allow users to take photos of the IoT devices from their smart phone cameras.  However, a fully wireless hand or ring based wearable integrated with a camera for interaction has not yet been realized due to  SWaP challenges.

CyclopsRing \cite{CyclopsRingEnablingWholeHandChanEtAl2015} uses a palm-facing wired RBG camera with a fish eye lens placed between the fingers to capture whole hand  gestures. To identify objects for interaction in the environment, CyclopsRing detects  tags~\cite{artags} placed on objects. The camera however is wired to a power source and hence the design do not consider the small size, weight and power (SWaP) constraints of designing an end-to-end wearable ring system, which can drastically change the design decisions. ThermalRing \cite{ThermalRingGestureTagZhangEtAl2020}  uses a  wired infrared (IR) camera  to capture IR energy emitted from the wearer's  opposite hand to track the hand motion and perform gesture input. For object detection and identification, ThermalRing requires objects to be retrofitted with IR reflective patches called ThermalTags similar to the tags used in in CyclopsRing and elsewhere \cite{CBandFlexibleRingMiyaokuEtAl2007}. As before, ThermalRing uses wired cameras and does not address the SWaP constraints. {FingerReader and FingerReader2.0 \cite{10.1145/2702123.2702421, 10.1145/3264904} are wired camera rings designed for visually impaired users to read text or aid in shopping. FingerSight \cite{4371592} uses a camera-enabled ring for haptic sensing of distant objects, and TouchCam \cite{10.1145/3161416} combines IMUs and cameras for gesture support and body location classification. FingerReader, FingerSight, and TouchCam are all wired devices, tethered to a host machine or smartwatch. Unlike these works, we create a fully wireless, standalone, camera-enabled ring that meets the SWaP requirements expected in this form factor.}

EyeRing \cite{EyeRingEyeFingerNanayakkaraEtAl} designs a wireless camera-enabled ring for accessibility applications such as navigation, currency detection, and color detection. However, the prototype device is significantly large (around the size of a smart watch) and does not address the size and weight constraints for ring devices. Further, EyeRing does not support interaction and control of IoT devices.  Unlike all this previous work, IRIS does not require instrumenting existing objects in the environment with fiducial markers and can instead recognize objects using vision. {Contrary to conventional wisdom, we show that a fully wireless, camera-enabled, wearable ring that meets the SWaP requirements expected in this form factor is possible. Cameras consume significant power, and given the small battery capacities capable of fitting a ring form factor (<27mAh), our work introduces system-level optimizations necessary to integrate a camera into a ring without being tethered to another device.}

\noindent{\bf Other sensors for IoT interaction.} SeleCon \cite{SeleConScalableIoTAEtAl2017}, WristQue~\cite{WristQuePersonalSensorMaytonEtAl2013}, and Minuet \cite{MinuetMultimodalInteractionKangEtAl2019a}  use ultra-wideband (UWB) equipped watches to enable pointing gesture detection for IoT device selection and control. SeleCon requires instrumenting each IoT device with a UWB radio which limits its scalability. WristQue \cite{WristQuePersonalSensorMaytonEtAl2013} and Minuet \cite{MinuetMultimodalInteractionKangEtAl2019a} require  UWB modules on the user and in the environment to determine the device selection from pointing gestures. Minuet \cite{MinuetMultimodalInteractionKangEtAl2019a} utilizes multi-modal interaction by also capturing the users' voice. Despite the appeal of these UWB enabled point-to-select wearables, all these prior prototypes are wired and/or do not demonstrate real-time operation, and hence do not address the power or latency requirements of end-to-end interactive ring devices. RingIoT \cite{RingIoTSmartRingDarbarEtAl} consists of a ring instrumented with an IMU, IR transmitter, capacitive sensor, and OLED display. This again uses a wired setup for both power and data transfer and hence does not consider the SWaP requirments in its design.  TRing \cite{TRingInstantCustomizableYoonEtAl2016} enables device interaction by sensing embedded magnets placed within everyday objects. Similar to \cite{RingIoTSmartRingDarbarEtAl}, this approach is limited as it requires instrumenting devices with specialized sensors. Furthermore, TRing cannot enable interaction from a distance as the magnetic field sensing is limited to the near-field. Magic Ring \cite{MagicRingSelfcontainedJingEtAl2013} has an  onboard radio and antenna to communicate with an intermediary device that translates the RF signals from the ring into infrared signals used to control the IoT devices and appliances in the home such as TVs or fans. Ring Zero \cite{RingZeroLets} is a commercial ring that proposes to use an onboard IMU to capture gestures and enable fine grained control over music volume and light brightness, however it is unclear how the device selection mechanism is implemented and whether it can enable ad-hoc control over any IoT device  without pairing.

\section{IRIS}

Here, we begin by outlining the requirements for our system design.  We then present our wireless ring hardware system and finally describe our real-time neural network pipeline.

%We start by laying out the requirements for our system design. Next, we provide an overview of how our network is designed to handle typical user scenarios in the smart home. As each user's home is different, we will then dive into how our design incorporates human-in-the-loop learning for adaptability across homes. Finally, we conclude by introducing our hardware -- a wearable ring designed for all-day use that brings the network's capabilities to your finger.

% \begin{figure}[t!]
% %\vskip -0.15in
% \centering
%     \includegraphics[width=0.49\textwidth]{figures/ring-hardware-diagram.png}
% \caption{{\bf TODO}}        
% \label{fig:system-hardware}
% \end{figure}

\subsection{System Requirements}

Our proposed interaction modality should be as ubiquitous as voice, while eliminating the need for lengthy  voice commands. The input modality should be as simple and swift as flicking a light switch or rotating a dial, ensuring a socially seamless experience. Finally, the system should require no modifications to the smart home devices as that would place an unnecessary burden on device manufacturers. We arrive at three core design principles:
\begin{enumerate}
    \item A simple interface as intuitive as flicking a switch
    \item Ubiquity comparable to voice interaction
    \item No modifications to existing  IoT devices
\end{enumerate}

% To address challenge (1), we look to create a point-and-click interaction model for controlling smart home devices. Integrating any sort of camera, QR code, or receiver on the end-device would break requirement (3), and thus the means of sensing must be present on the user. In conjunction with (2), the solution we arrived at was to create a context-aware wearable device that could adapt to any smart home device it detects. 
% Maruchi: I removed this because it seemed redundant ^

To address this, we design a wireless, vision-enabled smart ring. Achieving this requires us to address the following  challenges.

%\squishlist
\begin{itemize}
\item {\it Size, weight, and power requirements.} The overall weight and size of the device should be in a similar ballpark as a regular-sized ring. While shrinking the ring to fit within these constraints is challenging, we must also consider all-day battery life so that it does not need to be charged in the middle of the day.

\item {\it Real-time operation.} IRIS must be quick and responsive. An upper bound to not interrupt the user's flow of thought is around one second~\cite{NielsenResponseTimes}. So, we target the maximum end-to-end latency for IRIS to be one second, and benchmark it against voice assistant solutions. % like Siri in our results.

\item{\it Instance-based object detection.} An image may not necessarily provide the specificity of a voice command. For example, in a scenario where a user has a set of identical blinds in two different rooms, they could say, "lower the blinds in room A" or "lower the blinds in room B." IRIS instead must detect which instance a particular object is based on the surrounding visual context. 

%\item{\it Generalization and continuous improvement.} Finally, IRIS must generalize to  real world homes and lighting conditions. Seldom is a smart device moved once installed, and IRIS can capitalize on this opportunity by learning the full scene in which an object is located. By incorporating a human-in-the-loop model, users can add embeddings to optimize performance for their home. 

%\squishend
\end{itemize}

% This is an incredibly challenging problem unseen in the object detection community [TODO: Fact check? Is this true? We didn't find an immediate solution].

\begin{figure}[t!]
\centering
    \includegraphics[width=0.49\textwidth]{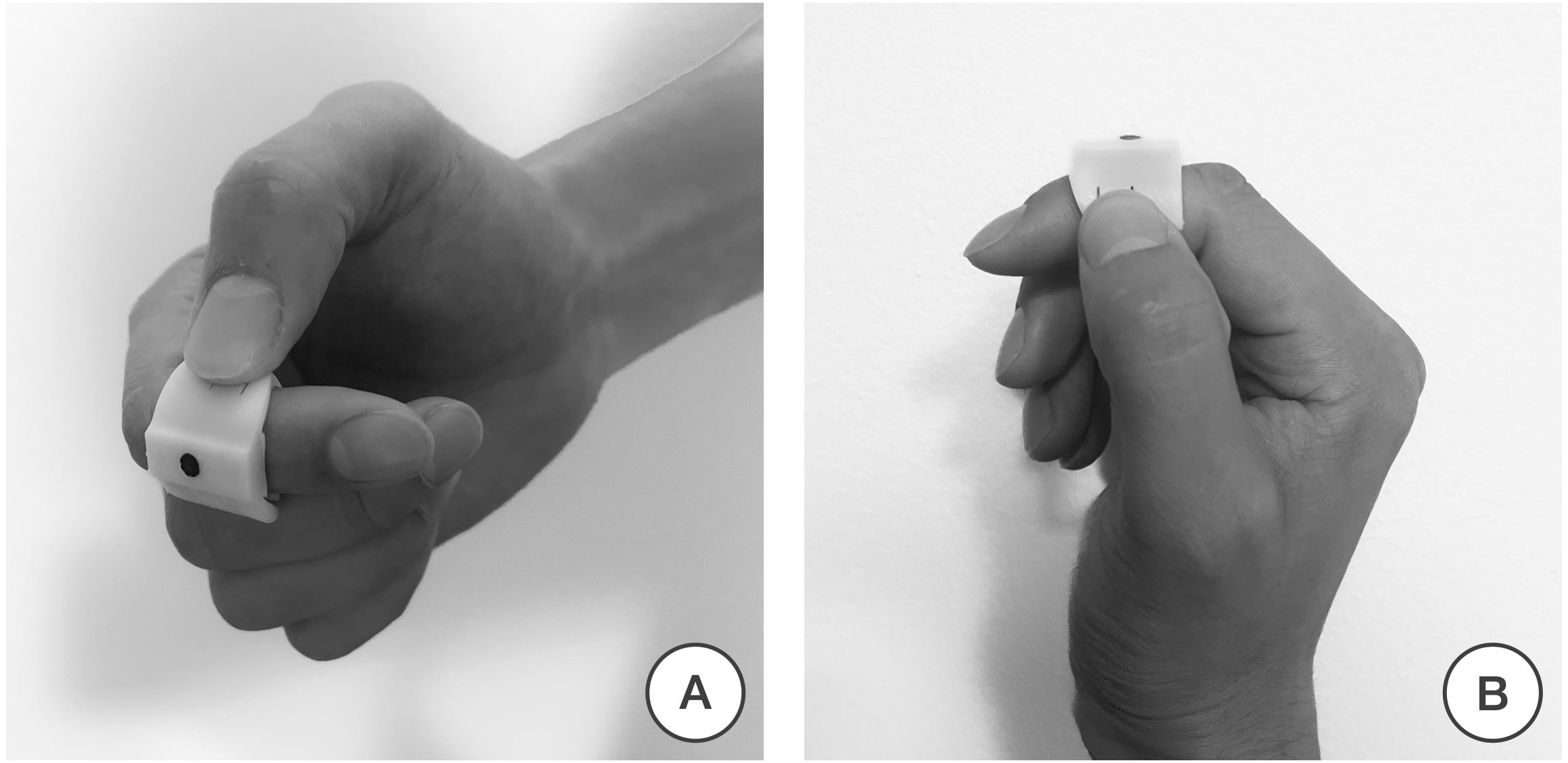}
\caption{{\bf IRIS on a User's Hand}. \textmd{(A) Front View, (B) Top View. }}
\Description{Two images (A) and (B) show IRIS worn on a person’s index finger with the black camera shown. Image (A) shows a front view of the ring on a person’s hand. Image (B) shows a top down view of the ring on a person’s hand.}
\label{fig:hand_model}
\vskip -0.15in
\end{figure}

\subsection{Wireless vision-enabled ring hardware} 
%We seek to stream images and IMU data from a wireless ring worn on the second digit of a user's index finger. This placement was chosen for two reasons: (1) guaranteed line of sight when issuing a gesture, and (2) allowing a user to "guide" the camera system with their thumb. The second was seen as a particular advantage over other configurations as this device has no feedback display like a traditional camera. Survey results (shown in a later section) also corroborate that the middle digit is easier to aim with than the proximal digit. Figure XXX, shows other placements considered during the design process. (TODO: Create table with images showing other placements showing which have LoS, which have a guiding mechanism, and which were most comfortable, which placement is most socially acceptable). Other core targets we aimed to achieve were a 24-hour battery life and a <500ms latency to provide the vision model with a reasonable amount of time for inference.

Here, we describe the various aspects of our wireless ring design.

\subsubsection{Hardware}
Our custom hardware design is comprised of an ultra-low-power 1/11" 320x320 QVGA CMOS image sensor (Himax HM01B0), a 6-axis inertial measurement unit (Bosch BMI270), and a Bluetooth Low Energy (BLE) microcontroller (Nordic nRF52840). The system is powered by a 27mAh battery and programmed via SWD over a Micro-USB connector. Our design is fully rechargeable through an on-board power management integrated circuit and provides the system with all necessary voltage rails (Maxim MAX77650). A single-pull single-throw (SPST) switch is used for gesture initiation, and images are streamed to a mobile phone for input into our neural network (see Fig.~\ref{fig:system-overview}).

\begin{figure*}[t!]
\vskip -0.1in
\centering
\includegraphics[width=1\linewidth]{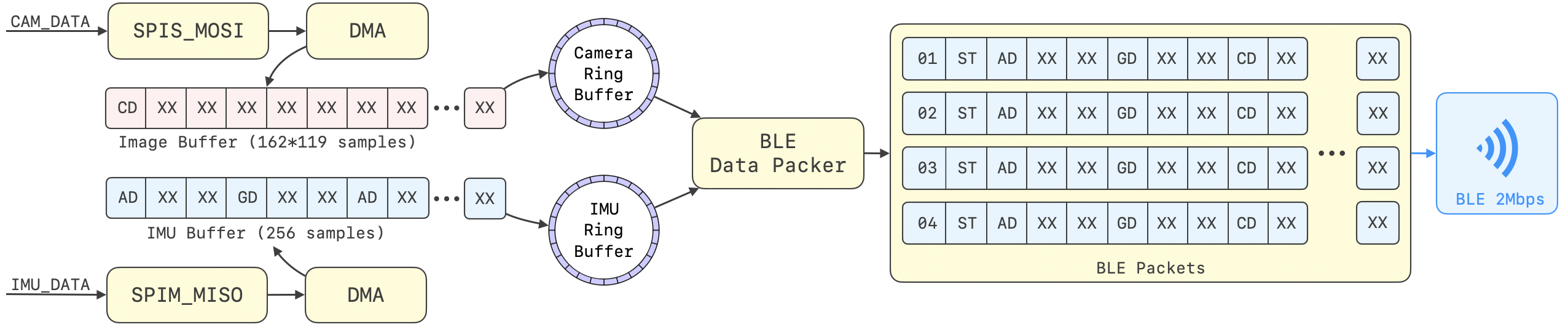}
\vskip -0.1in
\caption{IRIS image and IMU wireless data streaming.\textmd{ BLE packets are formed with the following structure: ST-Status Flags (Start of Frame, IMU Valid, Button State), AD-Accelerometer Data, GD-Gyroscope Data, CD-Camera Data.}}
\Description{A diagram showing the data streaming pipeline of the IRIS hardware and firmware. CAM_DATA (top left corner) goes to SPIS_MOSI block, which then goes to DMA and finally arrives at an Image Buffer of 162x119 samples (light red). Similarly, IMU_DATA (bottom left corner) goes to SPIM_MISO and then DMA to arrive at an IMU Buffer (light blue) containing 256 samples. Both light red and light blue buffer go into their respective ring buffers (purple circles). The data in these ring buffers get sent to BLE Data Packer, organizing it into BLE packets as shown in the large yellow box with light blue packets. These packets then get transmitted to the final light blue block on the right (BLE 2Mbps).}
\vskip -0.1in
\label{fig:data-flow}
\end{figure*}

\subsubsection{Wireless Latency} \label{subsec:wireless_latency}
The first challenge with our design is meeting a <500ms image acquisition latency target. IRIS utilizes the maximum supported  data rate of 2~Mbps over BLE \cite{nordicsemi_bluetooth_range}. To maximize throughput, we utilize the shortest connection interval available to iOS (15~ms) while transmitting 4 packets per interval (maximum supported by iOS) \cite{punchthrough_ble_throughput}. With a packet size of 247 bytes the effective data rate is 526,933 bits per second. The effective BLE throughput can be written as: 
\begin{equation*}
\text{{Throughput}} = \frac{{1000\,\text{{ms}} \times \text{{packets\_per\_interval}} \times \text{{packet\_size}} \times 8}}{{\text{{connection\_interval (ms)}}}}
\end{equation*}

The challenge is that a full 320x320 image is 819,200 bits, translating to 1562.5ms of latency (0.64fps) for the full image to be transmitted to the phone. So, we need to reduce the image size while preserving as much information as possible. To do this, first, we enable the QVGA window readout on the image sensor, which reduces the resolution to 320x240. This is still insufficient, so we utilize pixel binning to improve image acquisition time. Pixel binning is the concept of combining electrical charges from multiple adjacent pixels into a single "superpixel" {\cite{s150714917, Li2012}. Pixel binning essentially increases the effective pixel size, providing an improvement in SNR (signal-to-noise ratio) and low light. Furthermore, pixel binning also enables faster frame rates. By combining pixel data, the sensor can read out information from a smaller number of "superpixels" compared to the original number of individual pixels. This reduced readout time effectively leads to faster frame rates.\footnote{In IRIS, the BLE throughput is 527kbps, while the HM01B0 camera can stream a 320x320 image at 60fps (6Mbps). Reducing the resolution down to 160x120 allows IRIS to transfer more total frames over time.} Thus, while binning reduces the resolution of our system by a factor of four to QQVGA (160x120), it offers two key advantages: (1) a 4x improvement in signal-to-noise ratio, and (2) a 4x increase in frame rate. In this configuration, our image resolution is now 160x120 bringing the frame rate to 3.43fps, or about 290ms of end-to-end latency. This meets our design goal of less than 500ms of latency with some margin to spare.

\subsubsection{CMOS image sensor integration} 
Since the BLE SoC (nRF52840) has no dedicated camera interface, we configure the CMOS sensor (HM01B0) to operate in a 1-bit data transfer mode. The HM01B0 can then effectively act as a SPI controller and pass data to the nRF52840 through a SPI  (SPIS) port. This is achieved through signalizing the start of frame with a rising edge of Frame Valid (FVLD), and receiving the image bits through Pixel Clock Out (PCLKO) and Data (D0). These map to CS, SCLK, and MOSI, respectively. The table below shows a clearer representation of the mapping between a 1-bit camera interface and SPI:

\begin{center}
\begin{tabular}{ |c|c| } 
\hline
1-Bit Camera Interface & Serial Port Interface (SPI) \\
\hline

Pixel Clock Out (PCLKO) & Serial Clock (SCLK) \\ 
Data (D0) & Master-Out-Slave-In (MOSI) \\ 
Frame Valid (FLVD)& Chip Select (CS) \\ 
\hline
\end{tabular}
\end{center}

Since the maximum SPI port clock frequency on the nRF52840 is 8MHz, we  limit the pixel clock frequency accordingly. We achieve this by inputting an 8MHz signal into the HM01B0's  clock (MCLK) pin from the nRF52480. This synchronizes the two chips together and also limits PCLKO to 8MHz. The final detail to make this bus work is to invert the FVLD signal. The nRF52840 only supports an active low CS line, whereas FVLD is an active high signal. We circumvent this by triggering an active-high interrupt from FVLD and loop an external GPIO back to the nRF52840 to trigger an active-low interrupt for the SPI port to start receiving data. This  could be optimized by incorporating a NOT gate between FVLD and CS~\cite{RoboticsPaper}.

\subsubsection{Low-power design}
IRIS  manages power consumption through its three distinct power states: SLEEP, IDLE, and ACTIVE. In the SLEEP state, IRIS conserves energy by deactivating all hardware components except for an internal timer within the nRF52840 chip. This timer periodically awakens IRIS to check for the presence of a home WiFi network by receiving data from the connected mobile phone. While away from the user's home WiFi network, IRIS remains in this energy-saving mode, transitioning to IDLE only when a home WiFi connection is established.

In the IDLE state, IRIS enables all power rails and readies all peripherals, but clock-gates the camera and suspends the IMU. This ensures responsiveness while minimizing energy consumption. IRIS exits the IDLE state with a single button press via a hardware interrupt, transitioning it into ACTIVE mode. During this mode of operation, IRIS continuously streams data such as button state, IMU data, and camera pixels to the connected mobile device. After 3 seconds of streaming and inactivity from the button, IRIS exits ACTIVE and returns to IDLE to optimize low-power performance.

\begin{figure*}[t!]
%\vskip -0.1in
\centering
\includegraphics[width=1\linewidth]{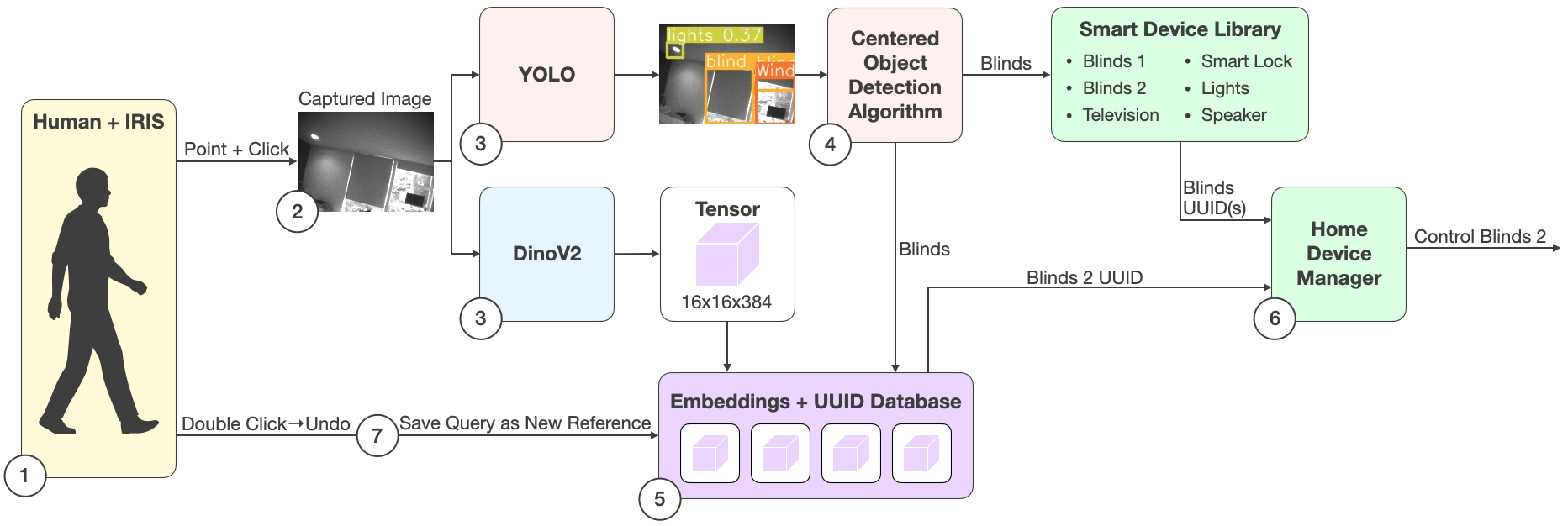}
\vskip -0.1in
\caption{IRIS pipeline. \textmd{(1) A user points and clicks at the smart device they would like to control. (2) IRIS wirelessly streams the  images to a smartphone, and (3) runs YOLO and DinoV2. (4) The centered object detection algorithm (CODA) filters out the multiple objects YOLO may detect and outputs the object closest to the center of the frame. IRIS then queries the smart device library, but since, in this example,  there are two instances of Blinds in the home it stops here and utilizes the Dinov2 path. Next, the output embedding from Dinov2 is passed as input to (5) the Embedding + UUID database to find the embedding with the highest similarity, and the output of CODA is also passed as input to reduce the search space. The highest similarity corresponds to Blinds 2 UUID, and the (6) Home Device Manager controls Blinds 2.  }}
\Description{Schematic of the IRIS neural network architecture. (1) shows a silhouette of a human and represents a user wearing IRIS. A point-and-click arrow then goes to (2) which shows a captured image from IRIS. (3) is a light red box showing YOLO, which outputs the same captured image from (2) with bounding boxes corresponding to objects it detects. Simultaneously, a light blue box (3) labeled DinoV2 outputs a Tensor of 16x16x384 which gets sent to (5). The output of (3) goes to (4) Centered Object Detection Algorithm which outputs Blinds to (5). Because there are 2 blinds in the Smart Device Library (light green box, top right), (5) outputs Blinds 2 UUID which corresponds to the 16x16x384 tensor. This gets sent to (6) Home Device Manager which controls Blinds 2. (7) shows a double-click gesture, corresponding to an undo operation. This allows users to save the most recent query as a new reference image in (5).}
\vskip -0.15in
\label{fig:ml-flow}
\end{figure*}

Our low-power design allows IRIS to operate continuously for 16-24 hours, depending on usage. Additionally, IRIS supports rapid charging capabilities, achieving a full recharge in just one hour with a 1C charge current, ensuring usage for extended periods.

\subsubsection{Fabrication}
The hardware schematic and layout for IRIS were designed using the open-source eCAD tool KiCad. A 2-layer flexible printed circuit was fabricated by PCBWay, while assembly was done by a local assembler. The 3D-printed enclosures were designed using AutoDesk Inventor and printed with a Formlabs Form 3 B printer using a liquid resin fabrication process. As illustrated in Fig \ref{fig:hand_model}, the camera sits behind the lid on the ring's outer surface, and the button is placed closer to the user's thumb while remaining horizontally aligned with the camera.

\subsubsection{Context-aware gestures}
We employ a simple gesture set that is adapted to the detected device. This set comprises two intuitive gestures: a single press for toggling device state and a press-and-hold with rotation, mimicking the action of a physical dial, for fine-grained control. Table.~\ref{tab:gesture_table}  illustrates how these two gestures can be used across an array of IoT devices.  A single press toggles the binary state of the device, while a press-and-hold with rotation allows for granular control. The rotation gesture was calculated by first deriving the tilt (in degrees) for the IMU axis that corresponded to rotation about the wrist from the accelerometer values using trigonometry. This tilt value was then offset by 90 degrees and scaled between 0 and 180 degrees. A running-difference of changes in this scaled tiled value was maintained and each incremental change in tilt was mapped to an increment change in volume or brightness. While the gesture set could be readily extended to additional devices such as garage doors, HVAC systems, and smart appliances, this work focuses on these five devices to maintain a manageable scope and data collection effort.

\begin{table}[h!]
\vspace{-0.05in}
\centering
\begin{tabular}{|c|c|c|}
\hline
\textbf{Device} & \textbf{Single Press} & \textbf{Rotation} \\
\hline
Lights & On/Off & Brightness \\
Speaker & Play/Pause & Volume \\
Smart Lock & Lock/Unlock & - \\
TV & On/Off & Volume \\
Blinds & Up/Down & - \\
\hline
\end{tabular}
\vspace{0.05in}
\caption{Context-aware gesture set.}
\Description{Table showing the context-aware gesture set for IRIS. The table shows how the same gesture (single press or rotation) will have a different effect depending on the device (first column). For example, a rotation for lights will change the brightness, but the same gesture can also adjust the volume of a speaker.}
\label{tab:gesture_table}
\vspace{-0.4in}
\end{table}

 %\textcolor{red}{OPTION TO REMOVE HERE SINCE WE DISCUSS IN FUTURE WORK? Furthermore, the potential exists for introducing more complex gestures, such as a press-hold-and-drag interaction for precise window blind positioning. However, we believe our current gesture set effectively exemplifies the core design principle of IRIS, with additional gestures left for future work.}

\subsection{Neural Network Pipeline}\label{sec:nn}

Our  system was designed with several key requirements. First, we aimed for a baseline level of "out-of-the-box" functionality to minimize setup for end-users, addressing one of the main pain points associated with new devices \cite{wollner2012evaluation}. Secondly, the vision model needs to support instance-based object detection to differentiate between multiple smart devices of the same class within the same environment. %Should we change this? Finally, the system should operate improve with usage. %, and adapt to emerging smart device categories and devices. 

Our pipeline consists of a fused model comprised of YOLOv8~\cite{ultralytics_2024} and DinoV2~\cite{oquab2024dinov2}. A YOLOv8 model was fine-tuned on a custom dataset of 10 classes to control a set of five common smart home devices. A fine-tuned YOLO model enables us to address the first requirement of out-of-the-box functionality. However, we still need to address our  other requirements, and YOLO fails to distinguish between two instances of the same class and also doesn't improve without training on additional data. Thus, we augment our system with Dinov2 to address the shortcomings of YOLO. To distinguish between two instances of the same class, IRIS provides a scanning feature, which enables users to take pictures when they have multiple instances of the smart device in their home. We create an embedding database with these collected pictures, and users can map reference images to a specific instance of a smart home device. Finally, users can address specific failure scenarios by adding additional reference images into the embedding database.

\begin{figure*}[t!]
\vskip -0.1in
\centering
\includegraphics[width=1\linewidth]{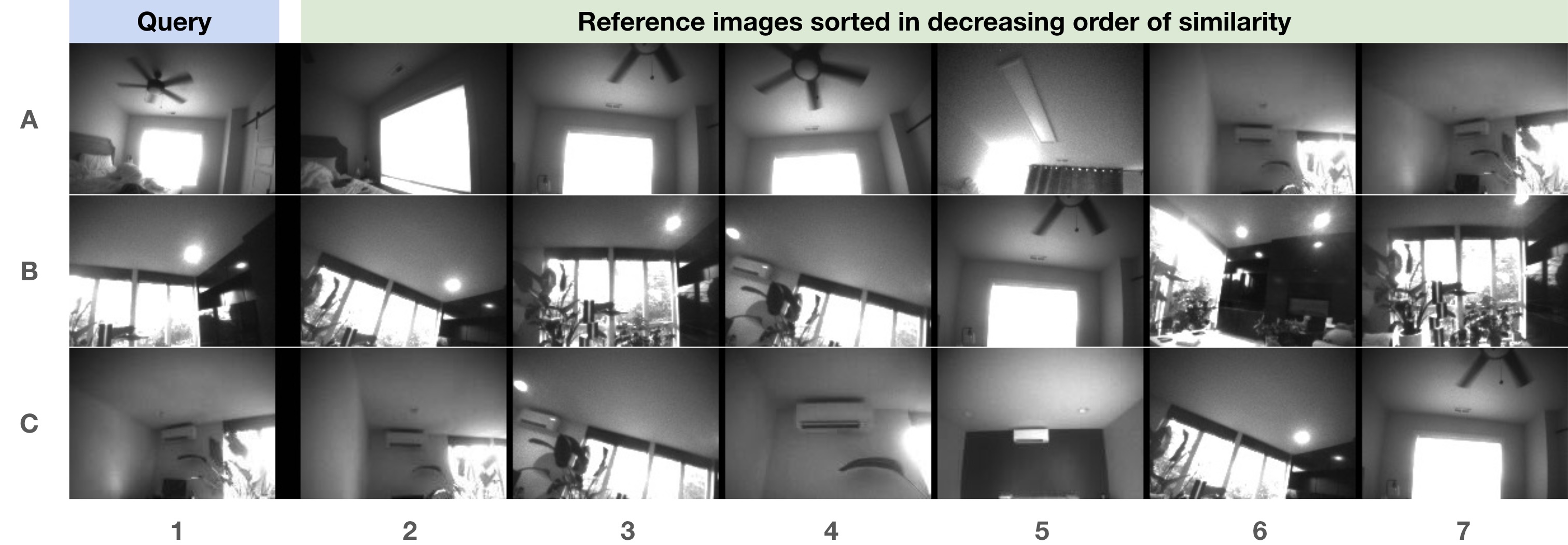}
\vskip -0.1in
\caption{{\bf{DINOV2 queries based on semantic similarity.} \textmd{Row (A) shows a query image of a blind (A1), while the maximum similarity reference image shows the same instance of the blind (A2). Row (B) shows a different instance of a blind (B1), and similarly shows how semantic similarity is used to find the reference image that represents the same instance (B2). Row (C) shows an image of an HVAC unit (C1), a class which is unlabeled in our YOLO dataset. Again, we observe that the same instance of the HVAC is returned (C2). A different instance of an HVAC (C5) further  shows that semantic scene understanding can be used to distinguish  between two instances of the same class.}}}
\Description{Grid of greyscale photos taken from IRIS ring showing a living room with a fan, window+blinds, and HVAC system on the wall. This figure shows how the DINOV2 query system works whereby query images (first column) are compared to reference images (rest of the columns) in the database. The reference images are ranked in decreasing order of similarity.}
\vskip -0.15in
\label{fig:dino-infer}
\end{figure*}

\subsubsection{Out-of-the box  object detection.}
Our primary focus was on object detection of five household objects: TVs, smart locks, blinds, lights, and speakers. We experimented with several well-known model architectures such as ResNet18~\cite{he_2016_deep_residual_learning}, ResNet50~\cite{he_2016_deep_residual_learning}, EfficientNetB0~\cite{tan_2019_efficientnet}, and YOLOv8~\cite{ultralytics_2024}. We moved forward with YOLO as it is an object detection model, allowing us to design a heuristic to control a specific device when multiple objects of interest are in the image. While YOLOv8 is trained on the COCO dataset~\cite{ultralytics_2024}, it lacked support for the five smart home devices we aimed to detect. To overcome this limitation, we collected over 2,000 images using a scraper on Google Images and approximately 500 images directly from IRIS's camera. The output of YOLO goes into an algorithm which iterates through each bounding box to retrieve the object that is closest to the center of the frame. This is achieved by calculating the Euclidean distance between the center of each bounding box with respect to the center of the image. This heuristic called CODA (Centered Object Detection Algorithm),  allows IRIS to return the object that is closest to what the user pointed at. 

%%Despite this, YOLO still has its shortcomings. With only YOLO, IRIS would fail to control a smart light if multiple smart lights are installed in one's home. While YOLO would output "lights", IRIS would be unable to identify which particular set of lights the user wants to control. Another limitation is that YOLO would fail when presented with a device that is not in our limited, hand-collected dataset. Examples range from a standard appliance backed by a WiFi-enabled smart plug, and when a new smart device is released. In the next section, we describe how DinoV2 is incorporated into our system, enabling IRIS to continue to function in these failure scenarios.
%Our approach for object detection and instance-based recognition draws inspiration from methodologies outlined in a paper published in Science Robotics~\cite{RoboticsPaper} \textcolor{red}{Is the insect paper pertinent to what we're saying here?}.

\subsubsection{{Human-in-the-loop learning with semantic similarity}}
\label{dino_method}
The above bounding box results combined with our heuristic disambiguate input frames with multiple objects, and provide us with a single object classification. This classification is sufficient in scenarios where there is one object (device) per class that the user wishes to control. However, if a user seeks to control multiple devices belonging to the same class (e.g., two TVs, one in the living room and the other in a bedroom), we would need to classify specific instances of the objects; beyond just object-level classification.  Taking this into account, in addition to our out-of-the-box solution based on a pre-defined set of classes, we implemented a human-in-the-loop model that allows users to define a separate class for each device.

%Furthermore, we need the ability for the users to determine what the ring should detect beyond pre-defined classes to support never-before-seen IoT devices. Specifically, YOLO alone will fail if presented with a device that is not in our limited, hand-collected dataset.

The intuition behind such a capability is that a device in a home is characterized by not just its own visual features but that of the surroundings it is placed in as well. For example, two smart speakers of the same make could be distinguished by the visual characteristics of the room it's placed in. This necessitates us to extract not just the features of the object of interest in the frame, but the features of the entire frame. We leverage  advances in large-scale self-supervised learning of visual features such as CLIP \cite{radford2021learning} and DINOv2 \cite{oquab2024dinov2} to compute semantic features at the image level. Specifically, we use the open-source DINOv2 \cite{oquab2024dinov2} model in our implementation, to obtain image-level semantic features. For a given image resized to 224x224 resolution, we split the image into 4x4 grid and compute a $R^{382}$ embedding for each patch in the grid. We note that the embedding corresponding to each patch considers not just the visual features within the patch but their relation to the rest of the image patches as well.

The user would first perform a setup with the IRIS system by capturing a few (1-5) images for each device they wish to control. During this setup, the IRIS app computes a $R^{4 \times 4 \times 382}$ DINOv2 embedding for each reference image. For a smart home with $N$ connected devices placed at certain locations in the home, our system would create a database of a set of reference embeddings $\mathcal{X} = \{r_{ni} \in  R^{4 \times 4 \times 382} \ | \ n \in \{1,2,..,N\}; i \in \{1,..,5\}\}$ on the smartphone. Then during the regular use, the user points at the device they wish to control by pointing the IRIS ring at the device. This action leads the ring to capture the image of the scene  and send it to the IRIS app. Then, as shown in step 3 in Fig. \ref{fig:ml-flow}, the app computes a query embedding $q \in R^{4 \times 4 \times 382}$ corresponding to the frame captured by the user. In parallel, the YOLO model predicts which class the query belongs to. Based on this prediction, we compute a semantic similarity score between $q$ and each of the embedding belonging to the predicted class in $\mathcal{X}$ using the following algorithm:

\begin{lstlisting}[language=Python]
import torch
import torch.nn.functional as F
def compute_similarity(q: torch.Tensor,
                       r: torch.Tensor):
    q = q.view(-1, 382) # shape = [16,382]
    r = r.view(-1, 382) # shape = [16,382]
    scene_sim = 0.0
    for pe in q:
        # pe is patch embedding
        scene_sim += F.cosine_similarity(
            pe.unsqueeze(0), r, dim=-1
        ).max()
    scene_sim /= q.shape[0]
    return scene_sim
\end{lstlisting}

This step results in a set of similarities corresponding to each of the reference embeddings in $\mathcal{X}$. If $s_{ni}$ is the similarity corresponding to the $i$th reference image of $n$th device, then the device  that the user wants to control, $\hat{n}$, could be inferred with:
$$ \hat{n} = \argmax_n \ \{s_{ni} \in  [0.0, 1.0] \ | \ n \in \{1,2,..,N\}; i \in \{1,..,5\}\} $$
This approach is illustrated in  Fig.~\ref{fig:dino-infer}. On the left most column, we show query images that the ring captured during its regular use. The rest of the columns show different reference images captured during the setup. In each row, the reference images are sorted in decreasing order of similarity with the query image on the left. We can observe that the similarity accounts for not just the objects of interest, but also the surroundings the object is placed in.

\subsubsection{Latency reduction by combining the models.}  {A key challenge associated with embedding-based retrieval methods (like DINOV2), lies in the search time increasingly  linearly with database size. To address this, IRIS leverages the output from YOLO+CODA, which identifies the object closest to the image center. The output classification from CODA serves as a proxy for the user's point of focus, enabling us to reduce the search space within the user-collected reference images. This targeted search strategy noticeably  reduces the runtime required for DINO's query, enhancing end-to-end system latency without compromising accuracy. We evaluate this method across 86 reference images in \xref{sec:latency}}.

\subsection{Training Methodology}

% While YOLOv8 is trained on the COCO dataset, it lacked support for the five smart home devices we aimed to detect. To overcome this limitation
{While the COCO dataset YOLOv8 is trained on includes appliances in its set of 20 classes, it lacked support for the five smart home devices we aimed to detect. Furthermore, we do not require the detection model to predict bounding boxes for objects not in our five classes of interest. To address these aspects}, we fine-tune YOLOv8 by collecting pictures using a scraper on Google Images. While these web-scraped images were easy to collect, they fail to model our camera's characteristics like field-of-view, dynamic range, and low-light sensor noise. Thus, we also collected and created a dataset directly from IRIS's camera. We adopt a hybrid training methodology where we first train on our web-scraped dataset, and then fine-tune on real data from our hardware. Our results show that our network trained in this way generalizes to real-world images captured from IRIS's camera.

\subsubsection{Web-scraped data} \label{sec:synthetic-data} 
We collected over 2000 images of our desired classes. In total, this dataset contains 287 lights, 309 speakers, 521 smart locks, 178 door handles, 32 doors, 198 blinds, 197 televisions, and 210 windows from Google Images. We include doors and door handles as it is difficult for YOLO to accurately detect smart locks from far away due to their small size, ultimately resolving to few features from far distances. Additionally, being able to detect windows allow us to reject false positives (e.g. sunlight being recognized as a home light source). The system can fall back to detecting doors at far distances to ultimately trigger a smart lock. These images were then converted to greyscale and resized to 160x160. These transformations were done to mimic IRIS's camera specifications. Further, we applied data augmentations such as horizontal flipping, cropping (0-20\%), rotations (-15 deg, 15 deg), brightness (-15\%, 15\%), and exposure (-10\%, 10\%). These data augmentations bring the web-scraped dataset to a total of 3,526 images. We split this data into 90\% training and 10\% validation {subsets}.

\subsubsection{In-the-wild data} 
% While the output model from our web-scraped data was able to detect objects from our 5 desired classes correctly, it failed to operate on IRIS's camera.
{While the model trained on our web-scraped data was able to detect objects from our 5 desired classes correctly, it failed to adapt to image characteristics of IRIS's camera.} This is because the web-scraped images do not contain characteristics such as the camera's dynamic range, resolution, and sensor noise. To address this, we further fine-tune on a dataset of captured images from our ring. We collected 513 images across 5 different homes from various angles and lightning conditions to generalize to unseen environments. This dataset includes 98 blinds, 57 doors, 34 door handles, 228 lights, 24 smart locks, 73 speakers, 64 televisions, 37 windows, and 161 background (null) instances. Background images contain none of the objects that IRIS aims to detect; this is so that the model outputs fewer false positives. After data augmentations, our in-the-wild dataset comprised of 1225 total images. This dataset was split for 86\% training, 7\% validation, and 7\% testing {subsets}.

\subsection{Mobile Device Integration}

\begin{figure}[t!]
\centering
    \includegraphics[width=0.48\textwidth]{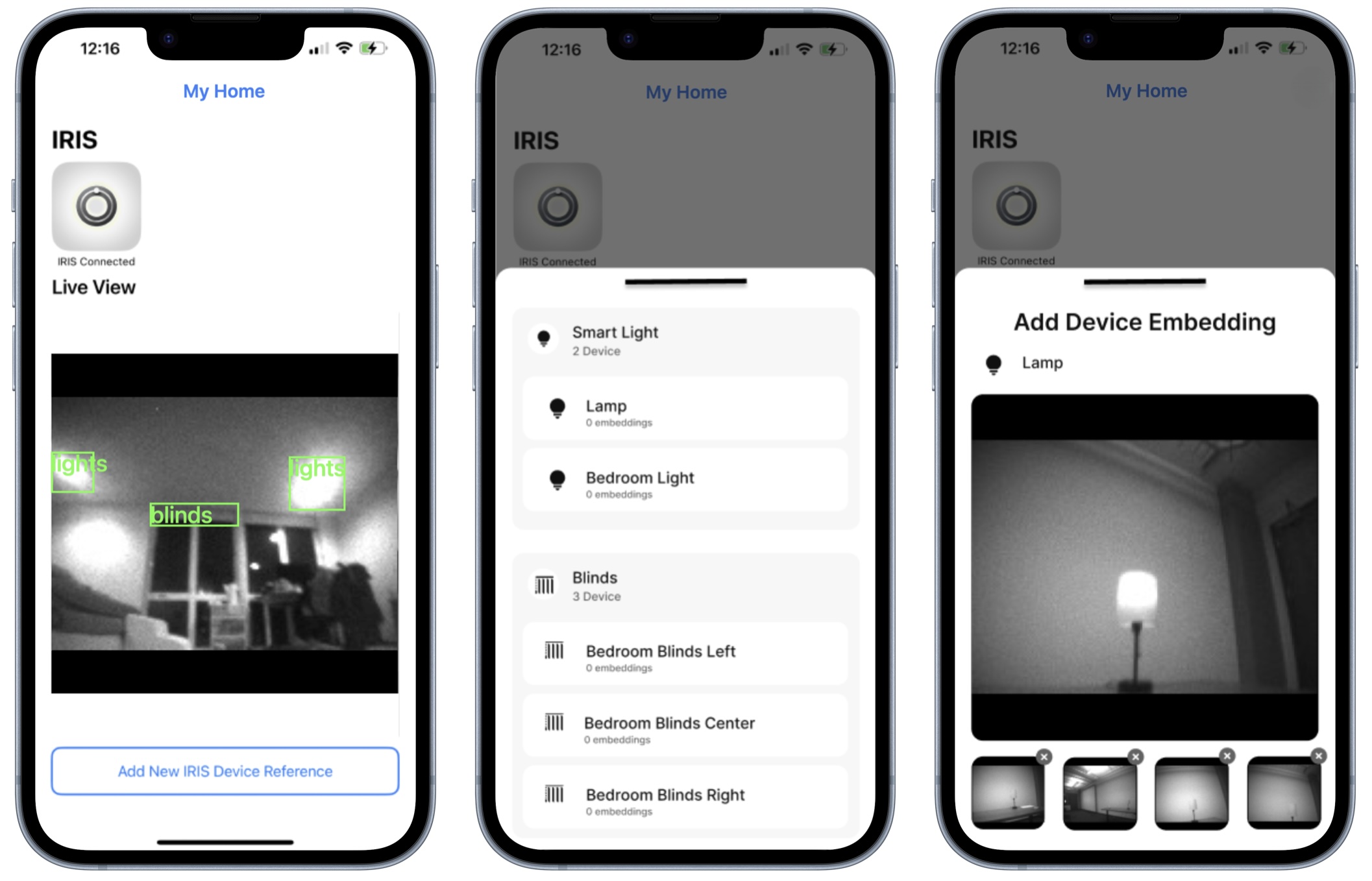}
\caption{{\bf{IRIS mobile app.} \textmd{(Left) Default view when user opens  the app. Users can observe the IRIS camera output; primarily for debugging and experimental use. (Middle) User can select a specific device to take new reference images for. (Right) User views the IRIS camera through the app to create new reference images.}}} 
\Description{Photo of three views of the user interface for the IRIS mobile application. A camera stream from the ring is shown of a set of blinds and two ceiling lights. The second view shows smart home devices listed in a database, for a user to add reference images for. The last view shows a view to Add Device Embedding with an image of a light. The four images below this image are other reference images associated with the light.}
\vskip -0.25in
\end{figure}
% HomeKit

Upon initial app launch and relevant user permissions, i.e., accessing control of smart home devices and Bluetooth, the HomeKit framework retrieves information on user-registered homes, smart home devices, their capabilities, configurations, and unique identifiers. Using this metadata, we build a smart device library and an embeddings+UUID database which our Home Device Manager uses to control the desired end device.

After IRIS connects to the iPhone, it waits for the button to be pressed prior to sending data wirelessly. Once a user presses the button to initiate a gesture, IRIS enters the ACTIVE state and starts streaming wireless data packets, which contain a sequence number, status flags, IMU vectors, and image pixels. {Each image frame is split and transmitted across multiple data packets, and} the sequence number is a monotonically increasing 8-bit value used to invalidate image frames in the event of dropped packets. Status flags include fields such as the button state and start-of-frame. Since IRIS continuously streams images, the start-of-frame flag marks {the starting packet of a new image frame and acts as the boundary between consecutive frames. The start-of-frame flag also signals the end of the previous frame to start inference.} The data pipeline is shown in Fig.~\ref{fig:data-flow}. 

% Inference (YOLO, create embedding)
% \textcolor{red}{Lucas, Eyoel to fill in 4 sentences of implementation detail.}
% We deploy YOLOv8 as a CoreML package for object detection on Apple devices. Its detections are fed into CODA to identify the central object. DINOv2, due to limitations with CoreML, leverages ONNX Runtime for cross-platform deployment. It directly takes raw image pixels to generate embeddings compared against stored references. To maximize results, we prioritize comparisons based on YOLOv8's classification, continuing to all references if a threshold isn't met.

Our fine-tuned YOLOv8 model's weights are transformed to a CoreML Package which we then load into Apple's VNCoreMLModel to perform requests and receive predictions and their bounds/bounding boxes. The raw pixel values of the streamed images are transformed into a bitmap image called CGImage and preprocessed into the desired shape requested by the model. CODA then finds the center object by searching for the bounding box closest to the image center, and returns this as the classified object. 

% Scan flow @Maruchi/@ProfShyam I slightly changed the flow of this ML integration to mobile section by talking about YOLO -> CODA -> Persistent Storage -> DINOv2 please feel free to revert the changes if confusing
For IRIS to detect between multiple instances of the same class, the system requires reference images (embeddings) of the device to be saved to persistent storage. We implement a "scan mode" through the app, during which the user can point-and-click at the target device and scene to create embeddings via DINOv2. Users can then associate these reference images with a target device in their home. Due to limitations of CoreML, we relied on ONNX Runtime for simplified cross-platform machine learning integration to host our DINOv2 model. With the image preprocessing baked into the DINOv2 ONNX model, we pass in the raw pixel values to create the embeddings which will then be automatically compared against stored references.

{Ultimately, the retrieved embedding with the highest similarity acts as a key, which is then used to fetch the unique identifier (UUID) of the targeted smart device within the user's home. After retrieving the UUID, we control the corresponding device via HomeKit.}

%To maximize the results from our two independent models, we prioritize comparisons based on YOLOv8's classification, continuing to all references if a threshold isn't met.

%\input{sections/4-qualitative_study-1}

\section{Evaluation}

\begin{figure}[t!]
\centering
    \includegraphics[width=0.48\textwidth]{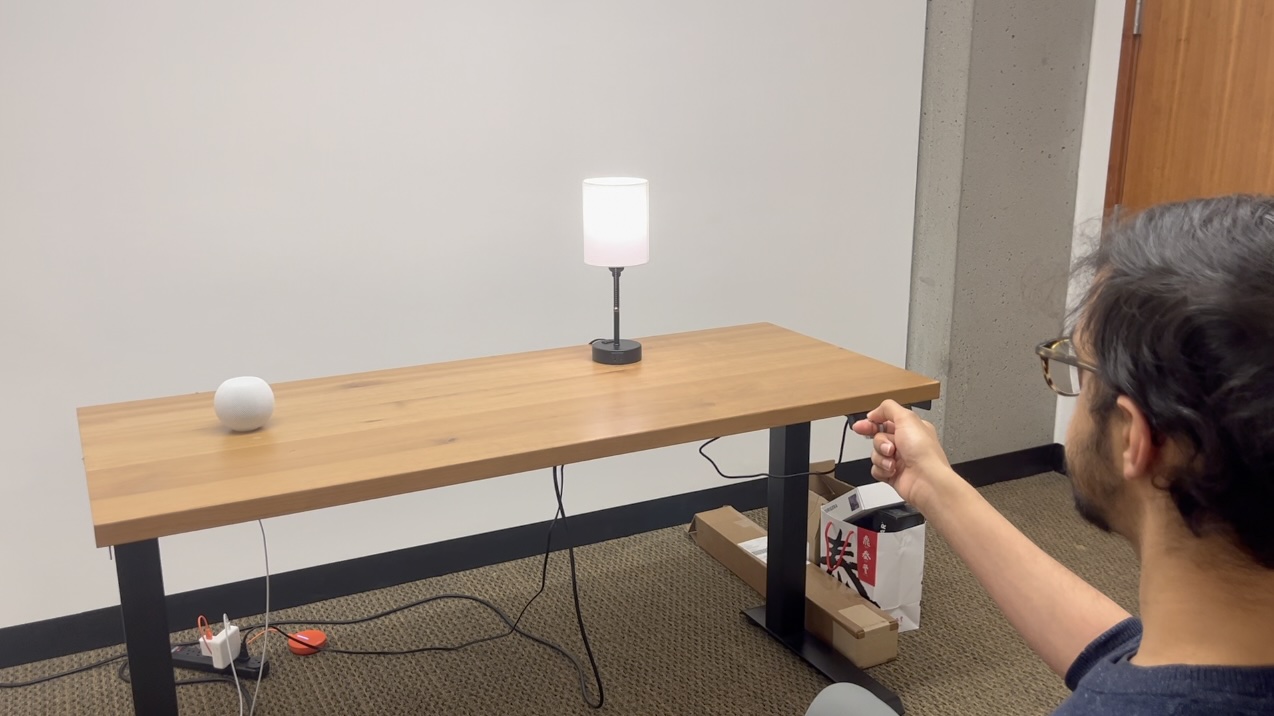}
\caption{{\bf{IRIS User Study.} \textmd{A participant from our user study interacts with a light and a speaker using our IRIS hardware.}}}    
\Description{Person is sitting in front of a smart speaker (left) and smart light (right). The person is pointing the IRIS ring at the smart lights and turning it on.}
\label{fig:user_study}
\vskip -0.2in
\end{figure}

\subsection{User Study}

% \begin{figure}[t!]
% \centering
%     \includegraphics[width=0.49\textwidth]{figures/placement.png}
% \caption{{\bf Mock smart-rings presented to  participants only in the first part of our study. \textmd{We solicit participants for their preferences between two  placements of a camera-based smart ring like IRIS: a ring worn on the proximal digit of the index finger (A) and a ring worn on the middle digit of the index finger (B). }}    }    
% \label{fig:placement}
% \vskip -0.1in
% \end{figure}

%Participants performed each gesture in our gesture set using these mock-rings — with markers showing a fake button and fake camera line of sight — and provided their ring position preference in terms of comfort and ease of aiming the camera.

We recruited 23 participants (16 male, 7 female) aged 18-35 {($\sigma$=4.72)} for a user study. Our study included an initial placement study to understand user preferences for the placement of IRIS along the index finger. The second part involved participants interacting in real-time with smart devices with IRIS and their voice. The participants were also provided a short questionnaire asking them about their overall experience and to compare IRIS against voice interaction. {Prior work \cite{10.1145/3484221} has shown that smartphone apps and screen-based interactions are even slower than voice control, and as such, were omitted from this study.}

\begin{table}[h!]
%\vspace{-0.1in}
\centering
\begin{tabular}{|c|c|c|c|}
\hline
\textbf{Question} & \textbf{Middle} & \textbf{Proximal} & \textbf{P-value} \\
\hline
Comfort & 52.2\% & 47.8\% & 0.339 \\
Targeting & \textbf{73.9\%} & 26.1\% & \textbf{0.005} \\
Preference & 43.5\% & 56.5\% & 0.202 \\
\hline
\end{tabular}
\vspace{0.05in}
\caption{Placement study results.}
\Description{Table showing the IRIS ring placement preferences of the participants in the user study. The table shows participant preference for comfort, ability to target devices, and overall preference for IRIS placement on either the middle digit or proximal digit. Participants were evenly split between middle and proximal placement for comfort (p value of 0.339>0.05). Participants preferred IRIS placement on the middle digit (p value of 0.005<0.05) for ability to target devices. Participants were also fairly evenly split between middle and proximal placement in terms of overall preference (p value of 0.202>0.05).}
\label{tab:placement_results}
\vspace{-0.3in}
\end{table}

\subsubsection{Placement study. }
Rings are typically worn on the proximal digit of a user's finger. However, conventional rings are not designed for aiming and controlling smart devices. While creating our hardware, we hypothesized that it would be easier to aim at devices and have more range of motion in one's hand if IRIS were placed on the middle digit of the index finger. {We provide participants with a set of metal rings (similar to IRIS's weight) for sizing their middle and proximal digits. Two markers on the rings indicate camera and button positions.}

% With this in mind, we provide participants with a set of rings to find two sizes that fit on the middle digit and proximal digit. 

% These two placements can be observed in Fig.~\ref{fig:placement}.

Once two rings were chosen, we asked participants to imagine that these mock rings were smart rings designed to control smart devices. We informed participants of the gesture set (point-and-click, point-hold-rotate, double-click), and had them perform these gestures on a lamp and a speaker. After evaluating both placements, they were provided with a short qualitative questionnaire.

The question set included: (1) \textit{Which configuration felt more comfortable for use?}, (2) \textit{Which configuration did you have an easier time "aiming" the camera with?}, and (3) \textit{Between the two configurations, which did you prefer?}. The responses to these questions are shown in Table \ref{tab:placement_results}. For each question, we conducted a one-tailed binomial test to understand if the result was statistically significant. Overall comfort was comparable between the two configurations with basically an even split across the 23 participants (52.2\%/47.8\%). While, 56.5\% of participants preferred the proximal position due to convention, this result isn't statistically significant enough to design the ring in this position. Ease of targeting the smart device between the two configurations was the one statistically significant result of this study where 73.9\% of participants preferred the middle digit. This result corresponds to a p-value of 0.005.

\begin{figure}[t!]
\vskip -0.1in
\centering
    \includegraphics[width=0.49\textwidth]{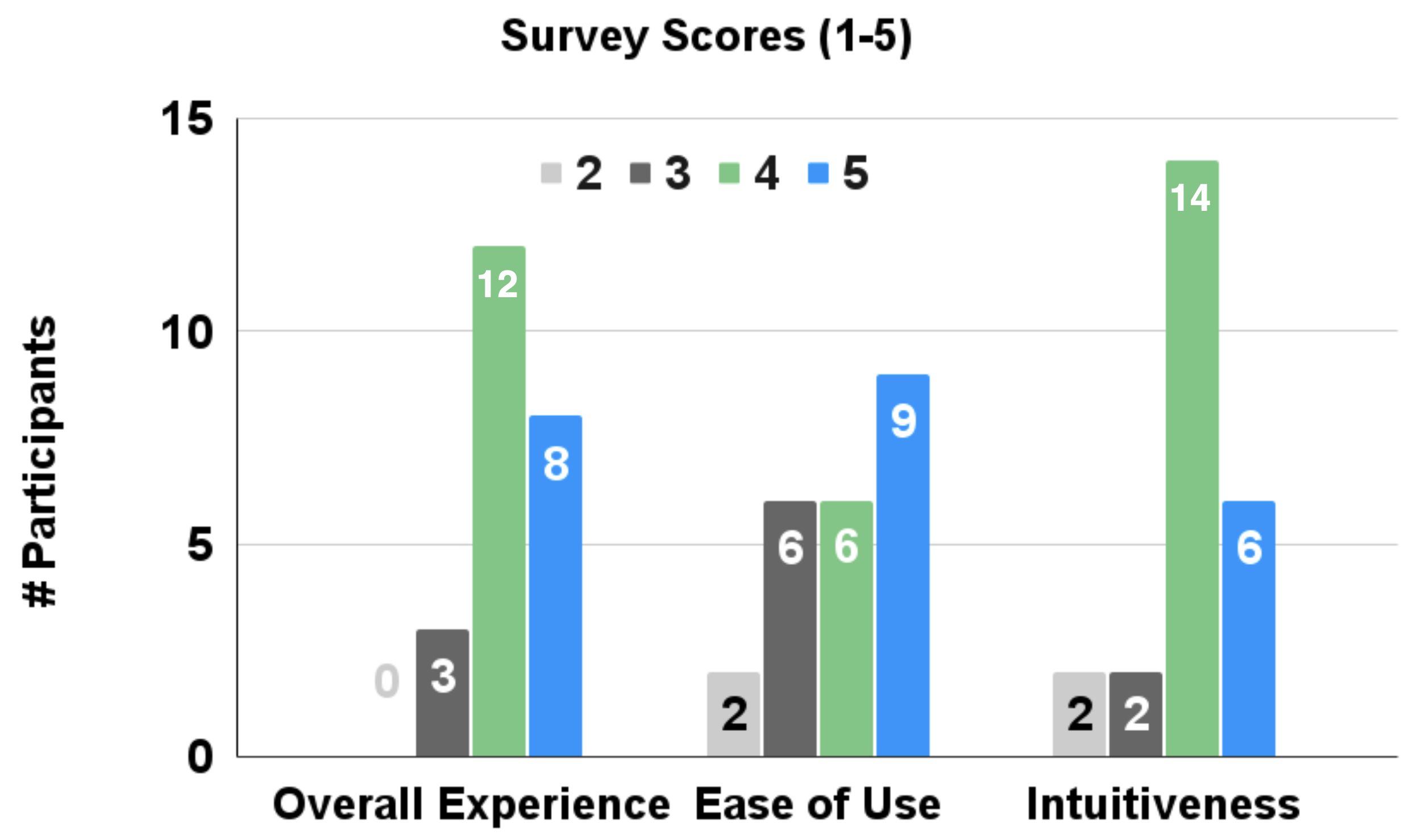}
    \vskip -0.14in
\caption{{\bf User interaction with our IRIS hardware. \textmd{Y-axis shows number of participants who ranked using a score from 1-5. Scores for 1 have been omitted as no participants responded with 1.}}}.      
\Description{Histogram of the user survey scores for using IRIS evaluated in terms of overall experience, ease of use, and intuitiveness. Participants ranked each metric on a scale of one to five with five being the best. Scores for 1 have been omitted as no participants voted 1. A majority of participants (12) gave the overall experience a four, followed by 8 participants giving the overall experience a 5. A majority of participants (9) gave IRIS a 5 for ease of use, followed by 6 participants giving a score of 4 and another 6 giving a score of 3 for ease of use. Lastly, a majority of participants (14) gave a score of 4 for intuitiveness, followed by 6 participants giving IRIS a score of 5 for intuitiveness. }
\label{fig:mos}
\vskip -0.25in
\end{figure}

\subsubsection{Interaction using IRIS hardware and voice.}
The goal of the next part of our study was to compare IRIS against a commercial voice assistant. As our neural network implementation and home device management were implemented on iOS, we opted to compare against Siri. Benchmarking against Siri guarantees that the communication under the hood (i.e., HomeKit) remains the same between the two interfaces.

This next phase of our study begins with introducing IRIS to the participants. To familiarize the participants with IRIS, we first play a short 30-second demonstration video. Next, we permitted them to try out IRIS by controlling a pair of smart devices, lights and speaker. After confirming their comfort and confidence using IRIS, participants engaged in a series of trials. The first trial involved a proctor who directed the participants to control either the lights or the speaker, toggling the device's state in response to cues. 10 of these trials were conducted for both IRIS and voice. {IRIS trials involved a hands-at-rest step to ensure that each event starts independently of the previous.} The second trial evaluated the granular control performance between IRIS and voice. Participants were told to set the speaker volume to that of a loud gathering or a pleasant ambient level based on if the proctor said "loud" or "quiet". In total, we collected 690 IRIS and voice trials.  Finally, these trials were all collected over video, which we use to assess the speed and efficiency between IRIS and a commercial voice assistant, later. 

% The demonstration video provided to users is available at:
%  \textcolor{blue}{{{\url{https://youtu.be/Bica-oG6Y08}}}}.

% A participant from our user study interacting with IRIS: \textcolor{blue}{\url{https://youtu.be/5zZzcOG2Hxk}

\begin{table}[t!]
%\vspace{-0.1in}
\centering
\begin{tabular}{|c|c|c|c|c|}
\hline
\textbf{} & \textbf{IRIS} & \textbf{Siri} & \textbf{No Diff} & \textbf{P-value} \\
\hline
Toggling State & 56.52\% & 39.13\% & 4.34\% & 0.202  \\
Granular Control & 78.26\% & 17.39\% & 4.34\% & \textbf{0.001}  \\
Social Acceptability & 69.57\% & 13.04\% & 17.39\% & \textbf{0.017}  \\
\hline
\textbf{} & \textbf{IRIS} & \textbf{Siri} & \textbf{Both} & \textbf{P-value} \\
\hline
Day-to-day Use & 34.78\% & 17.39\% & 47.82\% & 0.196  \\
\hline
\end{tabular}
\caption{Head-to-head subjective performance.}
\Description{Table showing the participant preferences for using IRIS versus Siri for toggling device state, granular control, social acceptability, and preference for day-to-day use. Participants showed no significant leaning towards using IRIS over SIRI for toggling device state (p value 0.202 is greater than 0.05) with a majority favoring IRIS. Participants significantly favored using IRIS over SIRI for granular control (p value of 0.001 less than 0.05) and in terms of social acceptability (p value of 0.017 less than 0.05). Lastly, participants showed no significant leaning towards day to day use between IRIS or SIRI (p value 0.196 greater than 0.05) with a majority of participants (47.8\%) indicating that they would use both day to day.}
\label{tab:head_to_head}
\vspace{-0.4in}
\end{table}

\subsubsection{Qualitative results.}

We conducted a qualitative analysis of the overall user experience, measured by Mean Opinion Score (MOS), and assessed subjective preferences between IRIS and voice among participants in our study. Participants were given 3 questions: 
\begin{enumerate}
    \item How was your overall experience using the smart ring? (1: Awful, 5: Amazing)
    \item How difficult was it to use the smart ring? (1: Very difficult, 5: Very easy)
    \item How natural was it to use the smart ring? (1: Very unnatural, 5: Very natural)
\end{enumerate}
%These questions were selected to address experience factors such as overall experience, ease of use, and intuitiveness. 
Across our participants, the mean opinion scores were 4.22, 3.96, and 4.17, respectively. The distribution of scores across each category is shown in Fig.~\ref{fig:mos}.  While these scores indicate a generally positive response, it is difficult to confirm without a direct comparison. To address this, we asked four specific questions regarding user experience between IRIS and a traditional voice assistant.
\begin{enumerate}
    \item Between using your voice and the smart ring, which would you prefer to use to toggle a device's state (i.e. turning a light ON or OFF)?
    \item Between using your voice and the smart ring, which would  you prefer to use to granularly control a device's output (i.e., a speaker's volume)?
    \item Between using your voice and the smart ring, which would you find more socially acceptable?
    \item Which would you be more inclined to use day-to-day?
\end{enumerate}
As shown in Table.~\ref{tab:head_to_head}, IRIS outperforms voice across all questions within our user study. Out of the participants surveyed, 13 expressed a preference for using IRIS for toggling the device's state, while 9 favored Siri. While the p-value is not statistically significant (0.202), we believe that this is attributed to the fact that our prototype's camera was angled slightly higher than it should have been. This caused some participants  to have to point lower than they expected, causing retries. Despite this, more than half of our participants surveyed still preferred IRIS to Siri. In terms of granular control and social acceptability, participants  preferred IRIS with p-values of 0.001 and 0.017, respectively. Considering that participants were using a prototype compared to a shipping commercial product, we believe this to be a notable result.

Finally, we surveyed participants and asked if they would prefer to use IRIS, Siri, or both for day-to-day use. 11 out of the 23 participants  surveyed mentioned that they would use both, and 8 participants  preferred IRIS, while 4 favored Siri. A p-value of 0.196 indicates that IRIS is not posed to replace voice assistants outright, but rather coexist alongside them. Participants who preferred to use both systems day-to-day were asked a follow-up question to explain further. Overall, they said that while controlling devices is easier and faster with the ring, there may be times when the ring is not worn, in which case the ubiquity of voice would be advantageous. Another popular comment was that there may be times (i.e., in a conversation) when using voice to control a device would be inappropriate as compared to the ring. A few participants highlighted how much easier it was to control volume with IRIS, and a single participant reported that voice control would be preferable if their hands were occupied and carrying other items.

\begin{figure}[t!]
    \begin{subfigure}[b]{.49\textwidth}
      \includegraphics[width=\textwidth]{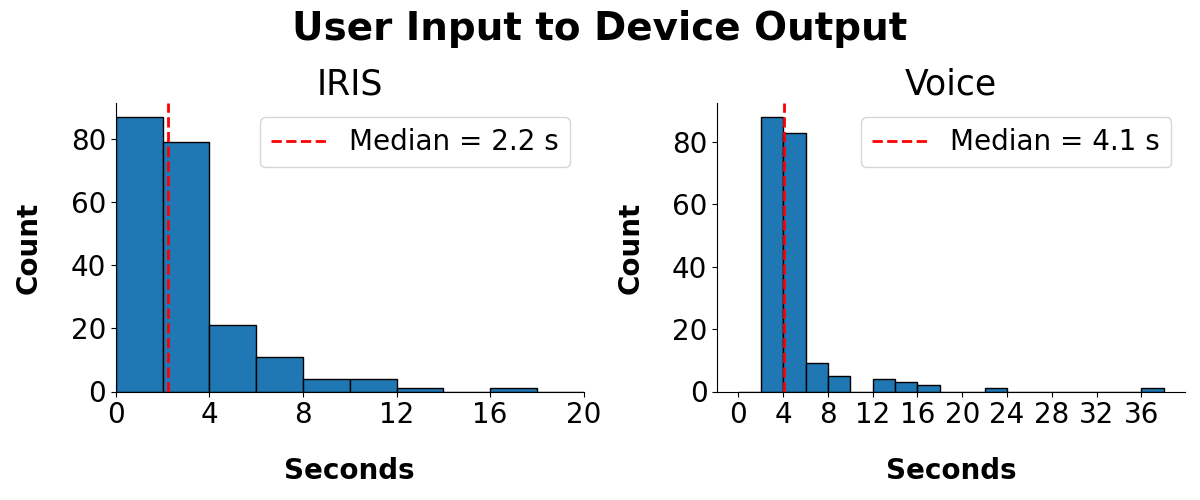}
      \vskip -0.23in
      \caption{}
    \end{subfigure}
    \begin{subfigure}[b]{.49\textwidth}
      \includegraphics[width=\textwidth]{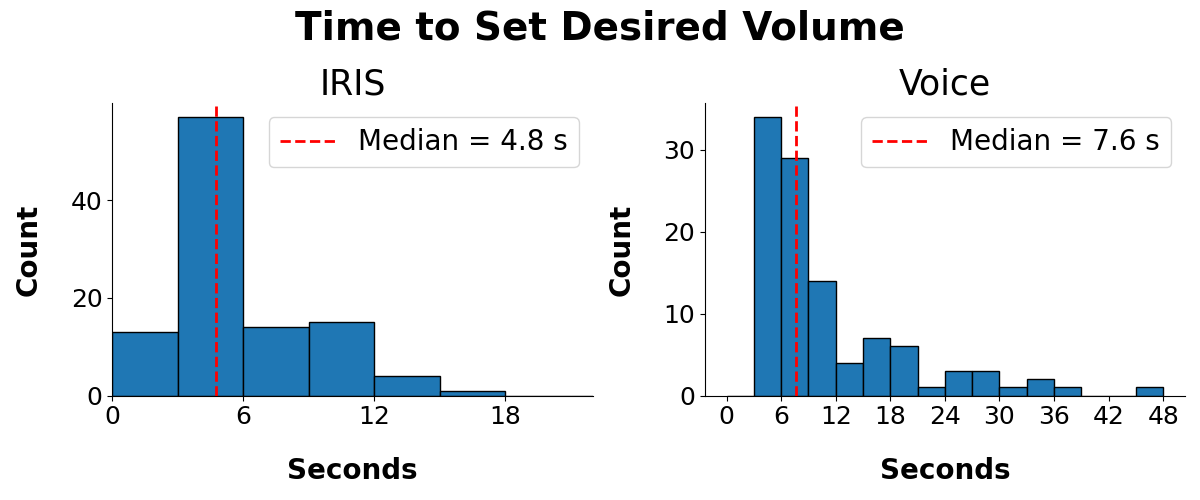}
      \vskip -0.23in
      \caption{}
    \end{subfigure}
    \vskip -0.1in
\caption{\textbf{Histogram of User Input to Device Output.} \textmd{ (UIDO) (a) and times to set desired volume (b) for IRIS and voice.}}
\Description{Four histograms are shown in this figure. The top two histograms show the user input to device output times across all participants in our user study (a). The bottom two histograms show time to set a desired volume across all participants in our user study (b). Both IRIS and SIRI show a positive skew for the user input to device output times. IRIS has a median user input to device output time of 2.2 seconds compared to voice with a median of 4.1 seconds. IRIS and SIRI also both show a positive skew for the time to set desired volume, with IRIS having a median time of 4.8 seconds compared to SIRI which has a median time of 7.6 seconds.}
\vskip -0.25in
\label{fig:user-study-histograms}
\end{figure}

\subsubsection{Quantitative results.}

% \begin{figure}[t!]
% \centering
%     \includegraphics[width=0.49\textwidth]{figures/tta_and_volume_no_outliers.png}
% \caption{{\bf User Input to Device Output and Time to Set Desired Volume boxplots.}}        
% \label{fig:tta-volume-boxplot}
% \vskip -0.2in
% \end{figure}

We conducted a quantitative analysis of the user input to device output (UIDO) time between IRIS and Siri for toggling the state (on/off) of a smart-light and smart-speaker as well as setting the volume for the smart-speaker. We measured the UIDO time for toggling the device state manually using a stopwatch. We started the timer at the end of each proctor's command issuance and ended it immediately after the light or speaker state changed. For the granular control, the end of the trial was determined by a visual confirmation from the participant that they had set the volume to a level that they deemed appropriate.

The histograms in Fig. \ref{fig:user-study-histograms} show the UIDO time for state toggling and volume control across all participant trials for IRIS and Siri control. Fig. \ref{fig:user-study-histograms}(a) shows that most participants were able to control the device state within 0 to 2 seconds using IRIS compared to 2 to 4 seconds when using Siri. Fig. \ref{fig:user-study-histograms}(b) shows that for granular volume control a majority of participants were able to set their desired volume within 3 to 6 seconds when using both IRIS and Siri. However, when using Siri there was a wider distribution of times, with a wider positive skew. 

% \begin{figure}[t!]
% \centering
%    \includegraphics[width=0.47\textwidth]{figures/angular-error.png}
%    \vskip -0.15in
% \caption{{\bf Angular error. \textmd{We attribute the higher vertical error due to the placement of IRIS's button being too far forward in respect to the camera. Users unintentionally tilted the camera upwards to press the button more comfortably.}}}        
% \label{fig:angle-error}
% \vskip -0.2in
% \end{figure}

For both control scenarios, the median UIDO and time to set desired volume was lower for IRIS (2.2 and 4.7 seconds respectively) than Siri (4.1 and 7.6 seconds respectively). This can be attributed to the fact that  it took longer for participants to utter a full voice command than perform the button press and rotation gestures.

Since IRIS is impacted by the pointing accuracy of the user, we also calculated the angular error between the center of the image frame and the center of the bounding box for the intended device. The angular error was calculated by dividing the field-of-view (FOV) of the camera (87 degrees) by the resolution, which was 160 to arrive at an angular-FOV/pixel of 0.54 degrees/pixel. We then multiplied this value by the x and y pixel errors calculated for 2344 photos gathered throughout the duration of the study. The median angular error was 7.6 degrees in the horizontal direction and 23.4 degrees in the y direction. We attribute the larger error in the vertical direction due to the relative position of the button and the camera on IRIS. In the current design, the button is positioned too far forward which required participants to have to rotate the camera towards them for their thumb to reach the button, resulting in the camera being angled higher vertically than intended.

\subsection{Vision Pipeline Evaluation}

\begin{figure}[t!]
%\vskip -0.1in
\centering
    \includegraphics[width=0.38\textwidth]{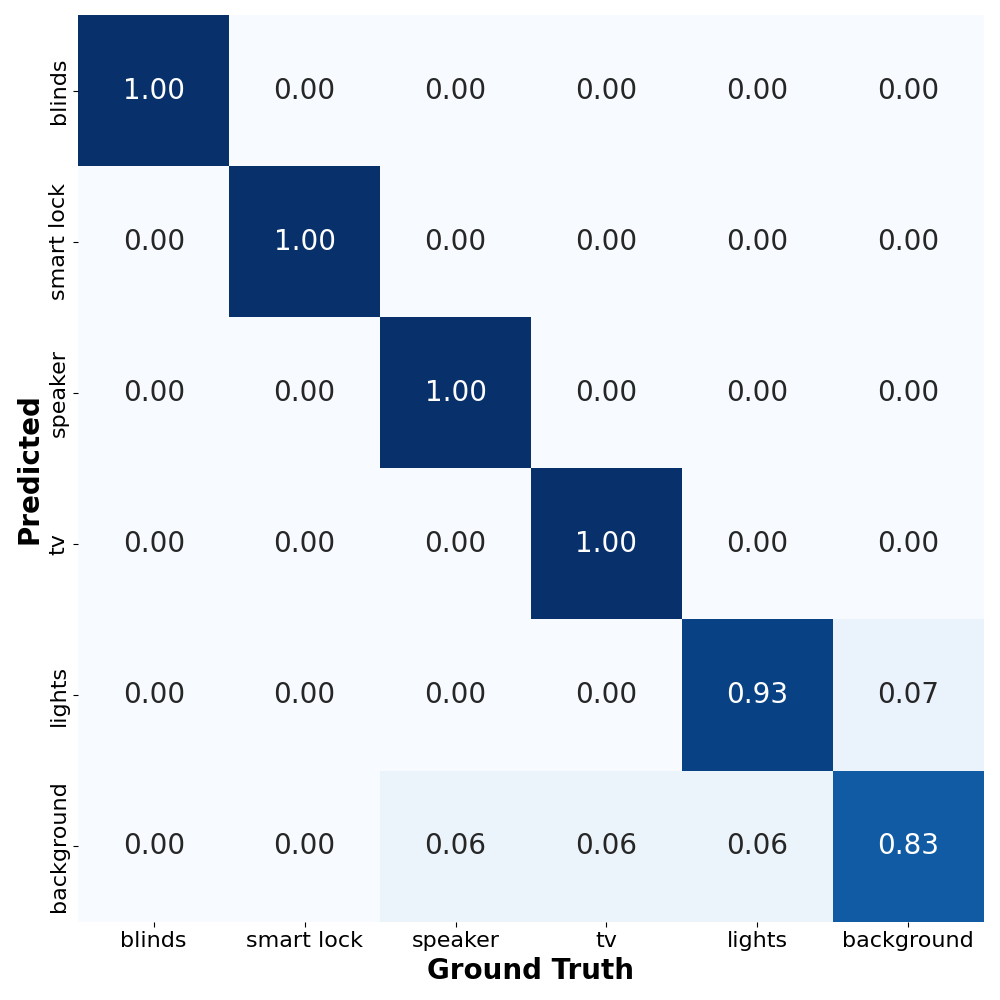}
\vskip -0.15in
\caption{{\bf Confusion matrix results across our five supported smart device classes and background. \textmd{We report a classification accuracy of 95.3\% after processing the image through CODA.}}    }    
\Description{Confusion matrix results across five supported smart device classes using YOLO (You only look once) and CODA (centered object detection algorithm). There is 100\% prediction accuracy for blinds, smart locks, speakers, and TVs, 93\% accuracy for lights, and 83\% accuracy for background. Background was predicted as speaker 6\% of the time, TV 6\% of the time, and lights 6\% of the time. Lights were predicted as background 7\% of the time.}
\vskip -0.2in
\label{fig:confusion}
\end{figure}

We first assess the performance of YOLO+CODA and DINOV2. These models were evaluated independently as their joint performance primarily  reduces runtime by reducing the search space in the DINOV2 (Embedding) database (see Fig. \ref{fig:ml-flow}). Classification accuracies of our models are independent of one another. We later evaluate the latency of jointly running the models (\xref{sec:latency}).

\subsubsection{Out-of-the-box object detection} 
We evaluate our object detection performance across 5 homes, and the 5 classes IRIS supports out-of-the-box (blinds, smart-lock, speaker, tv, lights) along with background (null). The objects were located in a variety of rooms (bedrooms, living rooms, kitchen, and office) and varying lighting conditions and angles. There was no overlap in images between our training and test datasets.

As IRIS's YOLO network is a fine-tuned model on an existing architecture, we validate its performance based on mean average prediction (mAP) values on the hardware dataset's test set. We used the existing YOLOv8 weights and fine-tuned it on both the web-scraped data and in-the-wild data for 75 epochs each. After completing the training against the in-the-wild dataset, the model achieved the following results: 0.793 Precision, 0.784 Recall, 0.81 mAP50, and 0.462 mAP50-95.

\begin{figure}[t!]
\centering
    \includegraphics[width=0.38\textwidth]{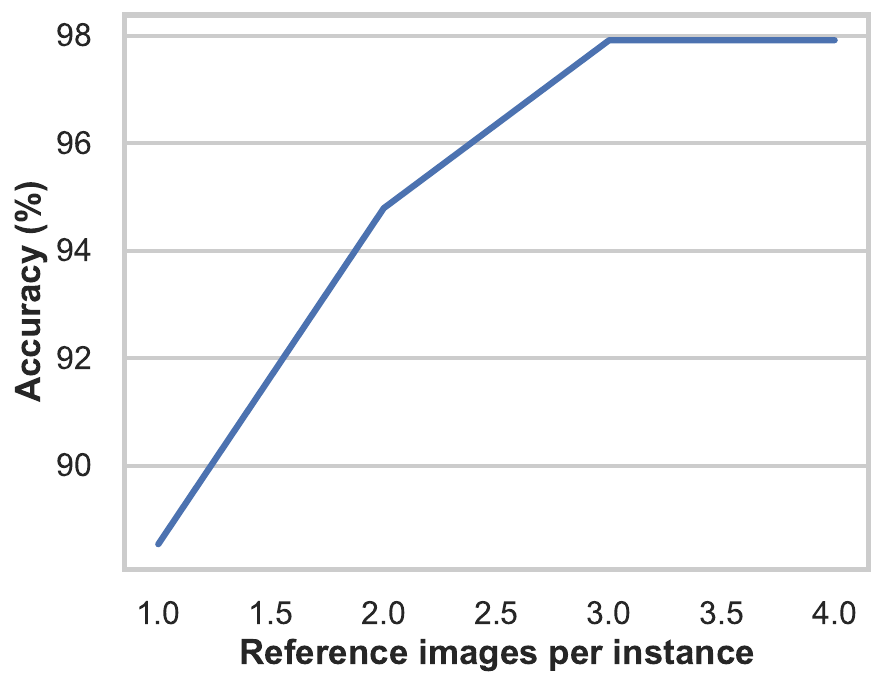}
    \vskip -0.15in
\caption{\textbf{ Accuracy of semantic similarity based search  as the number of reference images per instance increases. }. \textmd{{We report an accuracy of 98\% after 3 reference images.}}}
\Description{Line graph showing the accuracy of DINOV2 as the number of reference images for each device instance increases. For a single reference image in the database, DINO achieves an accuracy of approximately 88\%. For two reference images, the accuracy increases to about 95\%. For three images the accuracy increases to 98\% and plateaus for additional images beyond three. }
\label{fig:dino_sweep}
\vskip -0.15in
\end{figure}

\begin{figure*}[t!]    
\centering
\includegraphics[width=\textwidth]{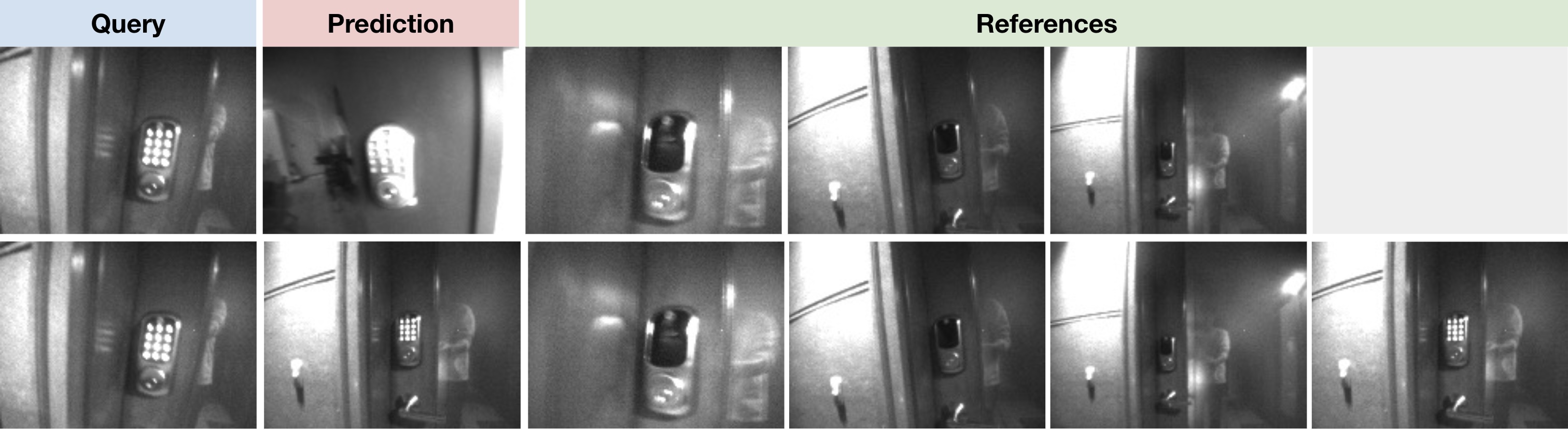}
\caption{{\bf {An example failure scenario with semantic similarity search.}    \textmd{This primarily occurs due to the lack of sufficiently diverse references. In the top row's inference with 3 reference images, the reference images corresponding to the correct instance do not contain the smartlock of interest in the on-state. So semantic similarity search wrongly picked a different smartlock instance in the on-state. In the bottom row, when we add a reference with the correct smartlock in the on-state as well, we pick the correct smartlock instance.}} }   
\Description{A two by six grid of photos captured from the IRIS ring showing different smart locks. In the first column, two images are shown of a smart lock with the label Query. In the second column, two images of smart locks are shown with the label Prediction. And in the last four columns, seven images of smart locks are shown with the label References. The bottom row shows that by adding an additional reference image, the query accurately predicts the correct instance of the smart lock, as opposed to the top row which shows a different instance of smart lock.}
\label{fig:dino_failures}
\vskip -0.1in
\end{figure*}

While the mAP50-95 score may not necessarily indicate strong overall performance, IRIS is dependent solely on the correct classification of the centered object. This is because if multiple detected objects exist in a frame, IRIS rejects the outlier objects and opts to control the object in the center of the frame. Thus, we computed a confusion matrix to show if the object in the center was classified correctly. Fig.~\ref{fig:confusion} shows the performance of our YOLOv8 model after CODA. Across 86 images, IRIS's YOLO+CODA implementation shows an accuracy of 95.3\% for center object classification. With only 3 false negatives and 1 false positive across the test set, we believe that the model performs sufficiently well for in-the-wild applications. We note that with a larger in-the-wild dataset, the performance of this part of our model could improve.

\subsubsection{Human-in-the-loop with semantic similarity}

{
For evaluating our instance-based detection method, we collect a user-defined test set of 96 images. These are all collected with the ring and contain 18 unique instances (devices): 2 blinds, 1 door, 4 lighting systems, 2 smart locks, 5 speakers, 2 TVs, and 2 HVACs. Each instance has 3-7 images associated with it taken from different perspectives. We sample one of these images \emph{without replacement} and use it as a query image. The rest of the images are considered references. We then predict the instance associated with the query image using the semantic similarity based search algorithm described in \xref{dino_method}. This prediction is compared against the ground-truth instance this query was sampled from, and marked as a correct or incorrect detection. This process is repeated across all 18 unique instances in the test set. Accuracy is computed over the entire set as the ratio of correct predictions over the total number of queries.}

To evaluate how many reference images a user needs to collect, we repeat this process while limiting the number of reference images per instance. This experiment allows us to gain insight by measuring the accuracy as a function of the number of reference images available per instance. The accuracy obtained from this evaluation is observable in Fig. \ref{fig:dino_sweep}. Our results indicate that 2 reference images are required per instance for 95\% accuracy and 3 images reaches an upper bound of 98\% accuracy. We note that these reference images are only required for when multiple instances of the same class are within the home and/or a new/unseen IoT device (e.g. HVAC) needs to be detected.

This approach results in very few inaccurate predictions (2\%), an example of which are shown in the Fig. \ref{fig:dino_failures}. Failures with semantic search primarily occur due to a lack of sufficiently diverse references. In the top row example, the model incorrectly predicts a different smart lock as the best match. We can observe that this is caused because none of the references contain the smart-lock in its \emph{on} state, but another reference of a different smart-lock contains an example in its on-state. We observe that when we include a reference with the queried smart-lock in its on-state, the prediction is accurate as shown in the bottom row of Fig. \ref{fig:dino_failures}.

\subsection{System Evaluation} \label{sec:endtoend}

\subsubsection{Latency.}\label{sec:latency}
End-to-end latency in our system is defined as the time elapsed between a participant completing a gesture (e.g. clicking the button) and the corresponding HomeKit command (e.g. toggling lights) being transmitted via an iPhone. This includes multiple components as shown in Table.~\ref{tab:latency}. We start our measurement by first putting IRIS into the IDLE state (camera clock-gated and IMU suspended) as this is the default state of IRIS when not in use. A hardware interrupt triggers upon image transfer completion through the nRF52840's BLE stack. We measured the hardware latency using an oscilloscope, calculating the delta between the button's falling edge and the image completion interrupt pin. This experiment yielded a hardware latency of 293ms, closely aligning with the expected value discussed in \xref{subsec:wireless_latency}.

\begin{table}[t!]
%\vspace{-0.1in}
\centering
\begin{tabular}{|c|c|c|c|}
\hline
\textbf{Embedding database size} & \textbf{4} & \textbf{50} & \textbf{100} \\
\hline
Hardware & 293ms & 293ms & 293ms \\
YOLOv8 & 28ms & 28ms & 28ms \\
Embedding Generation & 8ms & 8ms & 8ms \\
Embedding Query & 9ms & 244ms & 423ms \\
\hline
\textbf{Total latency} & 338ms & 573ms & 752ms \\
\hline
\end{tabular}
\vspace{0.05in}
\caption{\bf{Latency Breakdown.} \textmd{Query times are across the entire database (DINOV2 only) and are reduced via YOLO+CODA search space reduction. Results are shown in the figure below.}}
\Description{Table showing the latency breakdown across the hardware and neural network components as a function of the size of the DINOV2 embedding database size, with sizes of 4, 50, and 100 embeddings. The hardware latency is constant at 293 milliseconds at each database size.The Yolo classification latency is also constant at each database size, remaining constant at 28 milliseconds. And the embedding generation latency is also constant at 8 milliseconds at each database size. The embedding query latency increases from 9 milliseconds at a database size of 4 images, to 244 milliseconds for a database size of 50 images, to 752 milliseconds for a database size of 100 reference images. }
\label{tab:latency}
\vspace{-0.3in}
\end{table}

\begin{figure}[t!]
\centering
    \includegraphics[width=0.47\textwidth]{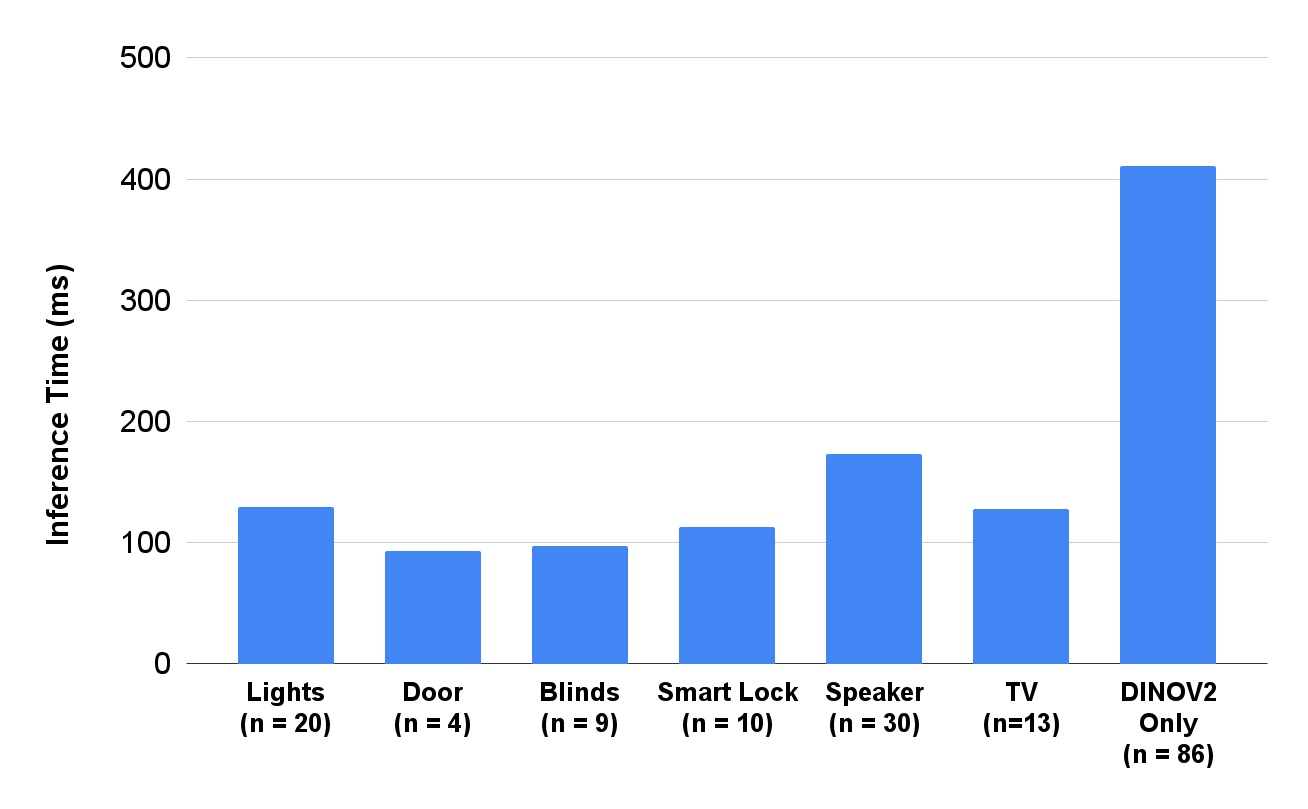}
    \vskip -0.15in
\caption{{\bf Query times for specific classes against the entire embedding database. \textmd{With only DINOV2, we report a runtime of 411ms. Utilizing the classification output from YOLO+CODA allows us to reduce the search space to a specific class, reducing latency by hundreds of milliseconds.}}}     
\Description{Bar chart showing the inference time for YOLO+CODA+DINO for a sample of lights (20 samples), doors (4 samples), blinds (9 samples), smart locks (10 samples), speakers (30 samples), and TV (13 samples), as well as inference time when using DINOV2 only (86 samples). The inference times for the individual classes are lower (less than 200ms) because we use the classification from YOLO+CODA to reduce the search space of DINOV2 Only (400+ms).}
\label{fig:yolo_dino}
\vskip -0.25in
\end{figure}

The other contributors to latency in our system are YOLO and DinoV2. We measured YOLO latency on the iPhone and observed values between 24-28ms. For Dinov2, we first measured the embedding computation time, which was 7.69ms. Next, we evaluated the time required to query the embedding database, noting that this latency scales linearly with the database size. We measured query times across a range of database sizes (4 to 100 embeddings), resulting in values ranging from 9ms to 423ms, respectively. Even in the scenario where IRIS queries 100 embeddings, the total system latency remains at 752ms. These measurements were all performed on an iPhone 13.

{To further optimize inference latency, we utilized the YOLO+ CODA output to reduce the search space for relevant embeddings that solely align with the classified device. We measure the runtime using this technique across an embedding database size of 86 images, corresponding to 16 unique device instances (2 blinds, 1 door, 4 lights, 2 smart locks, 5 speakers, and 2 TVs). We show in Fig. \ref{fig:yolo_dino}, that reducing the search space in this way reduces the query runtime on the order of hundreds of milliseconds. This latency optimization can be  quite significant for interactive mobile systems.} Finally, we note that further latency optimization could have been achieved by utilizing both a dedicated vectorized matrix library and parallelization. However, we leave this to future work.

\subsubsection{Power consumption}

\begin{table}
  \centering
  \begin{tabular}{c c c c}
    \hline
    \textbf{Component} & \textbf{SLEEP} &  \textbf{IDLE} & \textbf{ACTIVE}\\
    \hline
    SoC (ISP1807) & 4.23 $\mu$W & 3.53 mW & 19.2 mW \\
    IMU (BMI270) & 6.3 $\mu$W &  6.3 $\mu$W & 0.76 mW \\
    PMIC (MAX77650) & 23.5 $\mu$W & 23.5 $\mu$W  & 0.148 mW \\
    Camera (HM01B0) & -- & -- & 1.1 mW\\
    \hline
    \textbf{Estimated Total} & >34.03 $\mu$W& >3.76 mW& 21.2 mW\\
    \hline
    \textbf{Measured Total}  & 86.5 $\mu$W & 6.63 mW & 26.1 mW \\
    % Add more rows as needed
    \hline
  \end{tabular}
  \caption{\textmd{IRIS hardware power consumption.}}
  \Description{Table showing the IRIS hardware power consumption by component in each operating state: SLEEP, IDLE, and ACTIVE. The estimated total power consumption for SLEEP, IDLE, and ACTIVE are 34.03 microwatts, 3.76 milliwatts, and 21.2 milliwatts respectively. The measured total power consumption for SLEEP, IDLE, and ACTIVE are 86.5 microwatts, 6.63 milliwatts, and 26.1 milliwatts, respectively. }
  \vskip -0.2in
  \label{tab:power_consumption}
\end{table}

\begin{table}
  \centering
  \begin{tabular}{c c c}
    \hline
    \textbf{Gestures/hr} & \textbf{Battery Life} & \textbf{Battery Life  (SLEEP>8 hrs)} \\
    \hline
    10 & 16.6 hrs & 32.9 hrs\\
    30 & 15.9 hrs & 31.5 hrs \\
    60 & 14.9 hrs & 29.5 hrs \\
    \hline
  \end{tabular}
  \caption{\textmd{Battery life (hours) across different gesture rates for  the 27 mAh battery on our ring hardware.}}
  \Description{Table showing the IRIS battery life as a function of the number of gestures performed per hour. At 10 gestures per hour, the battery life is 16.6 hours for continuous usage and 32.9 hours for a full day's usage and 8 hours of SLEEP (corresponding to a user being away from Home WiFi for 8 hours).  At 30 gestures per hour, the battery life is 15.9 hours for continuous usage and 31.5 hours for a full day's usage and 8 hours of SLEEP.  At 60 gestures per hour, the battery life is 14.9 hours for continuous usage and 29.5 hours for a full day's usage and 8 hours of SLEEP. }
  \vskip -0.3in
  \label{tab:batt_life}
\end{table}

To calculate the expected battery life during usage, we first connected the battery terminals of IRIS to a power supply that has $\mu$A resolution. We measured the current draw of IRIS at 4.2V during each of the three power states (ACTIVE, IDLE, and SLEEP) for at least 25 seconds. For each operating mode, we averaged the current draw and derived the power consumption numbers shown in Table \ref{tab:power_consumption}. 

To calculate the expected battery life of IRIS, we began by assuming an average gesture duration of approximately 3 seconds in ACTIVE mode (26.1mW, 6.23mA). Over the course of a minute, this leaves 57 seconds for which the system is in its IDLE state (6.63mW, 1.58mA). From these measurements, we calculated the average power consumption for $N$ gestures over the course of an hour, giving an average current draw of 1.58mA for a single gesture  for one hour. Table \ref{tab:batt_life} shows the expected battery life of IRIS for different numbers of gestures performed per hour. Even at 60 gestures per hour or 1 gesture a minute, IRIS can last 15 hours on a full charge. Finally, we model IRIS's battery life for users who are away from home for up to 8 hours a day. This allows us to enter SLEEP mode (IRIS's lowest power state), as the user is away from their home Wi-Fi, and IRIS need not be constantly waiting for user input. In these use cases, IRIS is able to last beyond 24 hours.

{While the IRIS network consumes 774 MIPS during operation, this is exclusive to the ACTIVE state. Since IRIS's use case is episodic, the overall impact on a smartphone's battery life is minimal.}

% as \(\frac{{(6.23 \, \text{mA} \times 3n\, ) + (1.58 \, \text{mA} \times (3600-3n))}}{{3600 \, \text{s}}}\) which gives an average current draw of 1.58 mA for a single gesture ($n=1$) for an hour. Table  shows the expected battery life of IRIS for different numbers of gestures performed per hour. Even at 60 gestures per hour or one gesture a minute, IRIS can last for 15 hours on a full charge.

% \begin{table}
%   \centering
%   \begin{tabular}{c c }
%     \hline
%     \textbf{Power State} & \textbf{Current Draw (mA)}\\
%     \hline
%     ACTIVE & 6.23 \\
%     IDLE & 1.58 \\
%     SLEEP & 0.0206 \\
%     \hline
%   \end{tabular}
%   \caption{Measured current draw across operating states.}
%   \vskip -0.2in
%   \label{tab:current_draw}
% \end{table}

\subsubsection{Distance}

\begin{figure}[t!]
\centering
%\vskip -0.2in
    \includegraphics[width=0.47\textwidth]{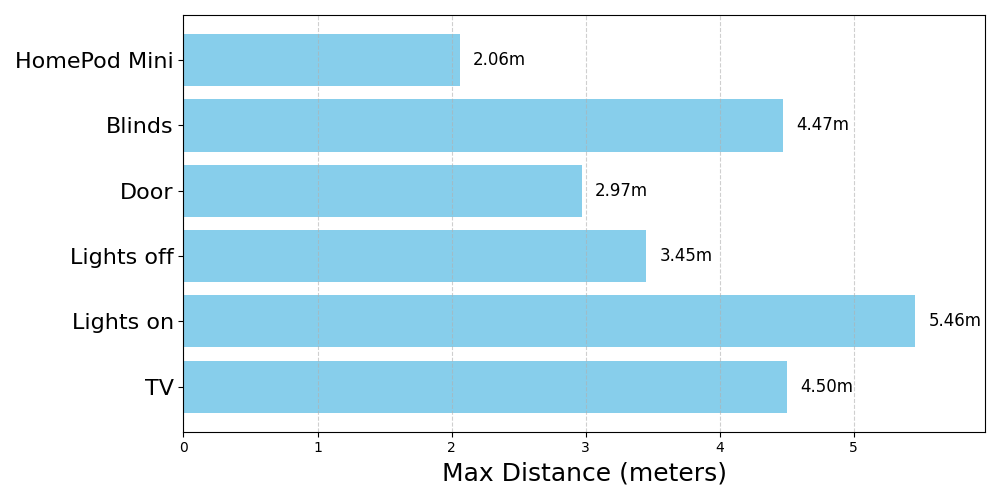}
    \vskip -0.15in
    \caption{{\textmd{ IRIS Distance Performance.}}}
    \Description{Bar chart showing IRIS classification performance as a function of distance to the smart device. The maximum distances at which IRIS can correctly detect and classify devices are 2.06 meters for the Homepod Mini, 4.47 meters for the blinds, 2.97 meters for the door, 3.45 meters for the lights when off, 5.46 meters for the lights when on, and 4.5 meters for the TV.}
\label{fig:distance}
\vskip -0.25in
\end{figure}

Our implementation enables us to observe IRIS's detection capability in real-time using an iPhone display. Thus, the effective detection range of IRIS was evaluated for various smart home devices through a visual feedback-based approach. For each device, we started at the minimum distance at which a bounding box was observed and progressively stepped backward. IRIS's detection capability at farther distances was evaluated based on the presence or absence of a bounding box around the target device. The furthest distance at which a bounding box was consistently generated for the device was recorded as the maximum detection range. Full detection range performance is observable in Fig.~\ref{fig:distance}.

This approach offered a practical way to assess IRIS's performance in a real-world setting. Our results aligned well with our expectations as small devices with few features failed first (HomePod Mini). Similarly, larger objects like blinds, televisions, and bright lights could still be picked up by YOLO at farther distances but fail once the object became small with respect to the image dimensions. The techniques described in \cite{ImprovedAlgorithmSmallObjectDetection} could potentially be used in future to improve performance further for small object detection.

\subsubsection{Low-light conditions.} We evaluate IRIS's object detection performance in low-light environments with a series of controlled experiments. We assessed IRIS's ability to detect smart home objects across a range of illuminance levels (lux). A controllable lighting system capable of adjusting brightness from 0 to 100\% was used to create a controlled testing environment during nighttime hours. We systematically decreased light intensity within the room until IRIS failed to detect the target device. The minimum lux at which a bounding box was consistently generated for each device was recorded as the minimum lux required for proper functionality.

Across the devices tested, we observed that IRIS detection failures began when the room's illuminance level (lux) approached or fell below 1. This indicates successful operation in quite dark environments. These results are largely due to two reasons. First, our camera is configured for auto-exposure, reducing its shutter speed in bright scenarios, while increasing its shutter speed in dark scenarios. {We note that the camera's maximum exposure time is 1/160th of a second, or 6.3ms. This is typically the upper bound for modern DSLR cameras, without in-body image stabilization, to be able to take images without motion blur from the user's hand.} Second, YOLO's feature extraction capabilities remain relatively strong until there is an absence of light. Overall, IRIS is able to maintain some level of detection for the target devices even in extremely low-light conditions. Passing and failing examples from this experiment are shown in Fig.~\ref{fig:low_light}.

\begin{figure}[t!]
\centering
    \includegraphics[width=0.47\textwidth]{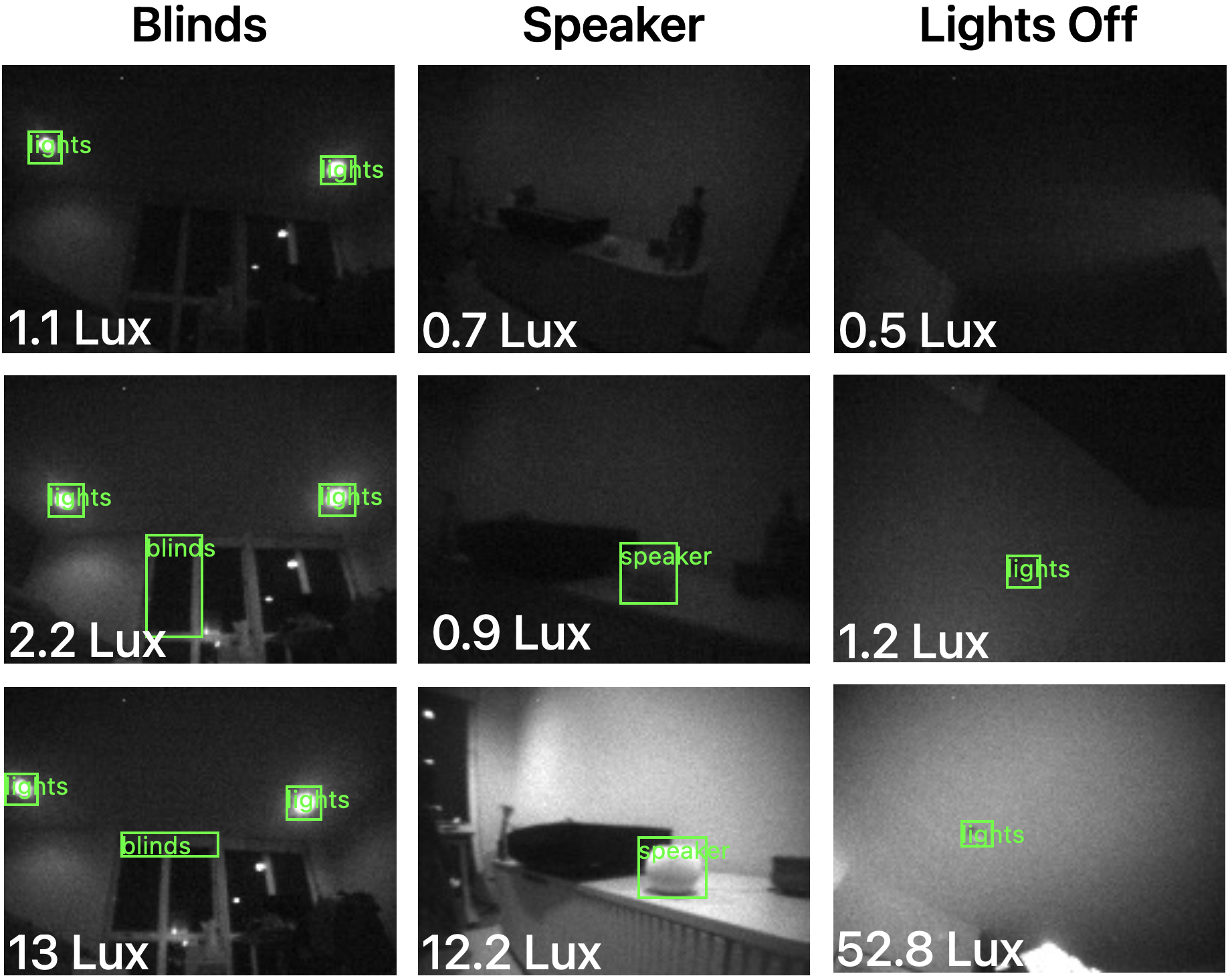}
\vskip -0.1in
\caption{{\bf Low-light testing examples. \textmd{Rows ordered in increasing illuminance. 50\% brightness set by our lighting system is approximately 12 lux, 20\%-2 lux, and 10\%-1 lux. Failures for speaker and lights occurred when our lighting system was turned to 0\%, while blinds failed at 10\%. All examples were collected at night.}}}
\Description{Six photos arranged in a three by three grid captured by IRIS. The column headings are of device categories. These are, from left to right, blinds, speaker and lights off. Each row shows a sample image of the column device at a given illumination level or lux. The illumination levels increase from top to bottom. The first row of photos are at low illumination and show a sample picture of a blind at 1.1 lux and 2 bounding boxes drawn around lights on the ceiling, a picture of a speaker at 0.7 lux with no bounding box, and a picture of lights at 0.5 lux with no bounding box. The second row is the same scene but at higher lux levels of 2.2, 0.9, and 1.2 lux for the blinds, speakers and lights off respectively. Each image shows a correctly drawn bounding box around the target device. The last row shows the same scene but at 13, 12.2, and 52.8 lux for the blinds, speaker and lights off respectively. All photos show correct bounding boxes drawn around each device. This figure shows how the auto exposure on IRIS assists the object detection in low lighting conditions.}
\label{fig:low_light}
\vspace*{-20pt}
\end{figure}

\section{Limitations and Discussion}

%Despite its approach, IRIS faces some limitations. Our system has been shown to struggle mildly in recognizing distance objects due to difficulty in resolving pertinent features. Additionally, users must be wearing the ring to utilize its benefits, and this might not be ideal for everyone. Furthermore, the ring's functionality is hampered when users' hands are full. While some limitations like reliance on a nearby phone for processing could be addressed by utilizing a local smart home hub for on-device inference,  the reliance on specific gestures for control could still be perceived as awkward by some users. These limitations highlight the need for further exploration to enhance user experience and broaden the ring's applicability.

%\subsection{Future Work}
Our work with IRIS lays the groundwork for several exciting future directions, which stem from the limitations of our current implementation. One area of exploration involves expanding the gesture set to incorporate more intuitive interactions {or additional utility}. {The results from the ring gesture elicitation study \cite{gheran2018gestures} show end users' preferences for an extensive referent gesture set performed with a ring to control various devices within a smart home. Many of these controls can be mapped to IRIS's simple gesture set, but additional gestures could enable IRIS to scale for devices that offer more functionality.} For instance, a press-hold-and-drag gesture could enable users to precisely control the stopping point for blinds. Similarly, sliding {or flicking} vertically and horizontally could be used to modulate light brightness or speaker volume, offering an alternative to rotation akin to a physical dial. {Adding redundancy through additional gestures would be beneficial for users to pick and choose gestures that feel most natural (i.e. IRIS could offer several gestures for controlling volume). Finally, GestuRING \cite{vatavu2021gesturing} presents hundreds of potential gestures for wearable rings, and in-air or on-body input could be utilized for controlling wearable devices, like headphones.}

{Detecting small objects across far distances (Fig.~\ref{fig:distance}) and/or unseen devices remains a challenge for our current implementation. In IRIS, users could take reference images of a desired device from farther distances, and the system would learn to utilize information from the scene to control the object. For unseen classes in our YOLO dataset, users could potentially take reference images of the unseen device and associate them with the device's corresponding UUID. While Fig.~\ref{fig:dino-infer}c shows the technically feasible of this for HVACs with our existing system, we did not test this extensively and leave this exploration to future work.}

Expanding IRIS's hardware presents another promising avenue for future research. Integrating a capacitive touch surface in place of, or alongside, the current button could enable users to navigate digital screens when pointing IRIS at the device. This could enable functionalities like navigating a television menu. Additionally, swiping across the touch surface could provide another alternative to rotations for scenarios requiring granular control. Finally, integrating a digital microphone could lead to a multimodal interface, and provide users the additional option of voice input. % alongside IRIS's vision based interface.

Running the neural network on the mobile phone affects its battery life. Offloading inference from the mobile device to the smart home hub is another promising avenue for improvement. This approach would benefit battery life by minimizing the computational demands placed on the user's phone. {Alternatively, embedding the inference on the ring could be a more privacy-preserving approach, alleviating the need to stream images wirelessly. With the recent advances in accelerator hardware, this could be a possibility in the near future.}

%Latency reduction is another area of focus for future optimization. Parallelizing the DinoV2 similarity queries presents a relatively straightforward approach to achieve this~\cite{dsds}. Furthermore, our current implementation utilizes non-vectorized matrix operations. Migrating to a proper vectorized matrix library has the potential to reduce latency further. 

%IRIS could also be baked into other interesting form factors. One such example is glasses with gaze tracking. Instead of pointing and clicking at a device with a wearable ring, users could instead look at the device they want to control, and pinch their fingers to toggle a device's state. This eye-tracking interaction technique was proposed in Gaze+Gesture \cite{gaze_gesture}.

%While IRIS was designed for smart home control, its core functionalities hold promise for broader applications. Here, we explore a few potential avenues for expanding IRIS's utility.

Since IRIS has limited use cases outside the home, incentivizing users to wear the ring all day is of notable importance. Integrating IRIS's features with existing health-tracking smart rings, such as Oura, Ultrahuman, or the Samsung Galaxy Ring \cite{ouraring, ultrahuman, forbes_samsung_galaxy_ring}, would provide users a reason to wear the ring all day. IRIS would then augment existing smart rings by providing smart home control in addition to tracking daily health metrics. Another application would be to control on-the-go accessories like Bluetooth speakers and headphones. An accessibility application of IRIS could be for individuals experiencing speech difficulties due to conditions like stuttering, apraxia, or dysarthria. IRIS could offer a valuable alternative to the conventional voice assistant for the speech-impaired population. Exploring this however is not in the scope of this paper.

%Another application of IRIS would be to embed the technology into a different form factor like smart glasses. With gaze and gesture tracking \cite{gaze_gesture}, users could then control devices by looking at them and performing a simple gesture, such as a finger pinch, to trigger an action. This approach eliminates the need for physical pointing and clicking, potentially enhancing user experience.

%\input{sections/7-conclusion-1}
\section{Conclusion}

We presented IRIS, the first wireless ring form-factor system for vision-based smart home interaction. To achieve this, we made multiple technical contributions, including a SWaP-optimized wireless hardware design that integrates a camera within a low-power, wireless ring form factor. Additionally, we achieved instance-level classification that leverages contextual scene semantics. We do this through a combination of YOLO for object detection and CODA for centered object selection. This significantly reduced the search space for the DINOV2 model, optimizing for runtime latency while not sacrificing instance classification performance. Finally, system-level optimizations ensured real-time responsiveness and extended battery life to 16-24 hours. Evaluations from our user study and experiments demonstrate IRIS's effectiveness and user preference over traditional voice control interfaces. We believe that this work represents an important step that can influence the landscape of ring-based human-computer interaction.

% IRIS presents a significant advancement in smart home interaction, overcoming the limitations of current voice assistant interfaces (social discomfort, verbalizing lengthy voice commands). This novel approach leverages a wireless, camera-enabled ring to perform instance-level detection and scene understanding on a mobile device.

% Our key contributions include:

% * **SWAP-optimized wireless camera ring:** This novel hardware design integrates a camera within a wireless, low-power, wearable ring.
% * **Instance-level classification:** By combining YOLO for object detection with CODA for centered object selection, we significantly reduce search space for the scene-aware DINOV2 model, enabling accurate device identification.
% * **Real-time and low-power operation:** System and network optimizations ensure responsiveness (under 1 second) and extended battery life (16-24 hours).

% Evaluations in our user study and experiments across diverse environments demonstrate IRIS's effectiveness and user preference over voice control modalities. With its real-time, intuitive interface and superior user experience (2 seconds faster than voice commands), IRIS has the potential to revolutionize smart home interaction. This work represents a crucial step towards a more natural and user-friendly interface for smart homes.

\section{Author Contributions}
\textbf{Manuscript Preparation:} MK and SG led; AG and BV led specific sections; YL and EG contributed. 
\textbf{System conceptualization:} MK and SG conceptualized the system. 
\textbf{Hardware and Firmware Design and Development:} MK led design and architecture; MK and AG jointly contributed to development. 
\textbf{Neural Network Design and Development:} MK and BV led design and architecture; MK led YOLO development; BV led DinoV2 development and EG contributed.
\textbf{Data Collection:} MK, AG, YL, EG, and AB jointly contributed. 
\textbf{Mobile Application Development:} MK and YL led; AG and EG contributed.
\textbf{Interaction Design:} MK led overall design. 
\textbf{User Study Design and Analysis:} MK and AG jointly led and contributed.
\textbf{System Evaluation:} MK led overall evaluation; AG led analysis from specific sections.

\begin{acks}
The  researchers were partly supported by the Moore Inventor Fellow award \#10617, NSF Graduate fellowship and a Thomas J. Cable Endowed Professorship. This work was facilitated through the use of computational, storage, and networking infrastructure provided by the HYAK Consortium at the University of Washington.
\end{acks}

\balance

\bibliographystyle{ACM-Reference-Format}
\bibliography{references}

%%% -*-BibTeX-*-
%%% Do NOT edit. File created by BibTeX with style
%%% ACM-Reference-Format-Journals [18-Jan-2012].

\begin{thebibliography}{49}

%%% ====================================================================
%%% NOTE TO THE USER: you can override these defaults by providing
%%% customized versions of any of these macros before the \bibliography
%%% command.  Each of them MUST provide its own final punctuation,
%%% except for \shownote{}, \showDOI{}, and \showURL{}.  The latter two
%%% do not use final punctuation, in order to avoid confusing it with
%%% the Web address.
%%%
%%% To suppress output of a particular field, define its macro to expand
%%% to an empty string, or better, \unskip, like this:
%%%
%%% \newcommand{\showDOI}[1]{\unskip}   % LaTeX syntax
%%%
%%% \def \showDOI #1{\unskip}           % plain TeX syntax
%%%
%%% ====================================================================

\ifx \showCODEN    \undefined \def \showCODEN     #1{\unskip}     \fi
\ifx \showDOI      \undefined \def \showDOI       #1{#1}\fi
\ifx \showISBNx    \undefined \def \showISBNx     #1{\unskip}     \fi
\ifx \showISBNxiii \undefined \def \showISBNxiii  #1{\unskip}     \fi
\ifx \showISSN     \undefined \def \showISSN      #1{\unskip}     \fi
\ifx \showLCCN     \undefined \def \showLCCN      #1{\unskip}     \fi
\ifx \shownote     \undefined \def \shownote      #1{#1}          \fi
\ifx \showarticletitle \undefined \def \showarticletitle #1{#1}   \fi
\ifx \showURL      \undefined \def \showURL       {\relax}        \fi
% The following commands are used for tagged output and should be
% invisible to TeX
\providecommand\bibfield[2]{#2}
\providecommand\bibinfo[2]{#2}
\providecommand\natexlab[1]{#1}
\providecommand\showeprint[2][]{arXiv:#2}

\bibitem[Alanwar et~al\mbox{.}(2016)]%
        {SeleConScalableIoTAEtAl2017}
\bibfield{author}{\bibinfo{person}{Amr Alanwar}, \bibinfo{person}{Moustafa Alzantot}, \bibinfo{person}{Bo-Jhang Ho}, \bibinfo{person}{Paul Martin}, {and} \bibinfo{person}{Mani Srivastava}.} \bibinfo{year}{2016}\natexlab{}.
\newblock \showarticletitle{SeleCon: Scalable IoT Device Selection and Control Using Hand Gestures}. In \bibinfo{booktitle}{\emph{Proceedings of the 10th ACM Conference on Embedded Systems for Energy-Efficient Buildings}}, Vol.~\bibinfo{volume}{2017}. \bibinfo{publisher}{{IoTDI 2017 (2017)}}, \bibinfo{address}{Log Angeles, CA, USA}, \bibinfo{pages}{107--114}.
\newblock
\urldef\tempurl%
\url{https://doi.org/10.1145/3054977.3054981}
\showDOI{\tempurl}
\showeprint[pmid]{29683151}


\bibitem[Amazon(2024)]%
        {AmazonAlexaVoice}
\bibfield{author}{\bibinfo{person}{Amazon}.} \bibinfo{year}{2024}\natexlab{}.
\newblock \bibinfo{booktitle}{\emph{Amazon {{Alexa Voice AI}} | {{Alexa Developer Official Site}}}}.
\newblock {Amazon Alexa}.
\newblock
\urldef\tempurl%
\url{https://developer.amazon.com/en-US/alexa}
\showURL{%
\tempurl}


\bibitem[Apple(2024)]%
        {HomePod}
\bibfield{author}{\bibinfo{person}{Apple}.} \bibinfo{year}{2024}\natexlab{}.
\newblock \bibinfo{title}{{HomePod}}.
\newblock \bibinfo{howpublished}{\url{https://www.apple.com/homepod/}}.
\newblock


\bibitem[Boldu et~al\mbox{.}(2018)]%
        {10.1145/3264904}
\bibfield{author}{\bibinfo{person}{Roger Boldu}, \bibinfo{person}{Alexandru Dancu}, \bibinfo{person}{Denys~J.C. Matthies}, \bibinfo{person}{Thisum Buddhika}, \bibinfo{person}{Shamane Siriwardhana}, {and} \bibinfo{person}{Suranga Nanayakkara}.} \bibinfo{year}{2018}\natexlab{}.
\newblock \showarticletitle{FingerReader2.0: Designing and Evaluating a Wearable Finger-Worn Camera to Assist People with Visual Impairments while Shopping}.
\newblock \bibinfo{journal}{\emph{Proc. ACM Interact. Mob. Wearable Ubiquitous Technol.}} \bibinfo{volume}{2}, \bibinfo{number}{3}, Article \bibinfo{articleno}{94} (\bibinfo{date}{sep} \bibinfo{year}{2018}), \bibinfo{numpages}{19}~pages.
\newblock
\urldef\tempurl%
\url{https://doi.org/10.1145/3264904}
\showDOI{\tempurl}


\bibitem[Chan et~al\mbox{.}(2015)]%
        {CyclopsRingEnablingWholeHandChanEtAl2015}
\bibfield{author}{\bibinfo{person}{Liwei Chan}, \bibinfo{person}{Yi-Ling Chen}, \bibinfo{person}{Chi-Hao Hsieh}, \bibinfo{person}{Rong-Hao Liang}, {and} \bibinfo{person}{Bing-Yu Chen}.} \bibinfo{year}{2015}\natexlab{}.
\newblock \showarticletitle{CyclopsRing: Enabling Whole-Hand and Context-Aware Interactions Through a Fisheye Ring}. In \bibinfo{booktitle}{\emph{Proceedings of the 28th Annual ACM Symposium on User Interface Software \& Technology}} (Charlotte, NC, USA) \emph{(\bibinfo{series}{UIST '15})}. \bibinfo{publisher}{Association for Computing Machinery}, \bibinfo{address}{New York, NY, USA}, \bibinfo{pages}{549–556}.
\newblock
\showISBNx{9781450337793}
\urldef\tempurl%
\url{https://doi.org/10.1145/2807442.2807450}
\showDOI{\tempurl}


\bibitem[Chen et~al\mbox{.}(2018)]%
        {SnapLinkFastAccurateChenEtAl2018}
\bibfield{author}{\bibinfo{person}{Kaifei Chen}, \bibinfo{person}{Jonathan F\"{u}rst}, \bibinfo{person}{John Kolb}, \bibinfo{person}{Hyung-Sin Kim}, \bibinfo{person}{Xin Jin}, \bibinfo{person}{David~E. Culler}, {and} \bibinfo{person}{Randy~H. Katz}.} \bibinfo{year}{2018}\natexlab{}.
\newblock \showarticletitle{SnapLink: Fast and Accurate Vision-Based Appliance Control in Large Commercial Buildings}.
\newblock \bibinfo{journal}{\emph{Proc. ACM Interact. Mob. Wearable Ubiquitous Technol.}} \bibinfo{volume}{1}, \bibinfo{number}{4}, Article \bibinfo{articleno}{129} (\bibinfo{date}{jan} \bibinfo{year}{2018}), \bibinfo{numpages}{27}~pages.
\newblock
\urldef\tempurl%
\url{https://doi.org/10.1145/3161173}
\showDOI{\tempurl}


\bibitem[Darbar et~al\mbox{.}(2019)]%
        {RingIoTSmartRingDarbarEtAl}
\bibfield{author}{\bibinfo{person}{Rajkumar Darbar}, \bibinfo{person}{Mainak Choudhury}, {and} \bibinfo{person}{Vikalp Mullick}.} \bibinfo{year}{2019}\natexlab{}.
\newblock \bibinfo{title}{RingIoT: A Smart Ring Controlling Things in Physical Spaces}.
\newblock , \bibinfo{numpages}{2--9}~pages.
\newblock


\bibitem[de~Freitas et~al\mbox{.}(2016)]%
        {SnapToItUserInspiredPlatformdeFreitasEtAl2016}
\bibfield{author}{\bibinfo{person}{Adrian~A. de Freitas}, \bibinfo{person}{Michael Nebeling}, \bibinfo{person}{Xiang~'Anthony' Chen}, \bibinfo{person}{Junrui Yang}, \bibinfo{person}{Akshaye Shreenithi~Kirupa Karthikeyan~Ranithangam}, {and} \bibinfo{person}{Anind~K. Dey}.} \bibinfo{year}{2016}\natexlab{}.
\newblock \showarticletitle{Snap-To-It: A User-Inspired Platform for Opportunistic Device Interactions}. In \bibinfo{booktitle}{\emph{Proceedings of the 2016 CHI Conference on Human Factors in Computing Systems}} (San Jose, California, USA) \emph{(\bibinfo{series}{CHI '16})}. \bibinfo{publisher}{Association for Computing Machinery}, \bibinfo{address}{New York, NY, USA}, \bibinfo{pages}{5909–5920}.
\newblock
\showISBNx{9781450333627}
\urldef\tempurl%
\url{https://doi.org/10.1145/2858036.2858177}
\showDOI{\tempurl}


\bibitem[Dive(2024)]%
        {retaildive_2024}
\bibfield{author}{\bibinfo{person}{Retail Dive}.} \bibinfo{year}{2024}\natexlab{}.
\newblock \bibinfo{title}{27\% Increase in Smart Home Adoption Since 2020: YouGov Report}.
\newblock
\newblock
\urldef\tempurl%
\url{https://www2.deloitte.com/us/en/insights/industry/telecommunications/connectivity-mobile-trends-survey/2023/smart-home-industry-adoption-trend.html}
\showURL{%
\tempurl}


\bibitem[Dot(2022)]%
        {RingZeroLets}
\bibfield{author}{\bibinfo{person}{Daily Dot}.} \bibinfo{year}{2022}\natexlab{}.
\newblock \bibinfo{title}{Ring Zero: The Smart Logbar That Could Change How We Interact With Tech}.
\newblock
\newblock
\urldef\tempurl%
\url{https://www.dailydot.com/debug/ring-zero-smart-logbar-sxsw/}
\showURL{%
\tempurl}


\bibitem[Endo et~al\mbox{.}(1996)]%
        {endo1996using}
\bibfield{author}{\bibinfo{person}{Yasuhiro Endo}, \bibinfo{person}{Zheng Wang}, \bibinfo{person}{J~Bradley Chen}, {and} \bibinfo{person}{Margo~I Seltzer}.} \bibinfo{year}{1996}\natexlab{}.
\newblock \showarticletitle{Using latency to evaluate interactive system performance}.
\newblock \bibinfo{journal}{\emph{ACM SIGOPS Operating Systems Review}} \bibinfo{volume}{30}, \bibinfo{number}{si} (\bibinfo{year}{1996}), \bibinfo{pages}{185--199}.
\newblock


\bibitem[Fiala(2005)]%
        {artags}
\bibfield{author}{\bibinfo{person}{M. Fiala}.} \bibinfo{year}{2005}\natexlab{}.
\newblock \showarticletitle{ARTag, a fiducial marker system using digital techniques}. In \bibinfo{booktitle}{\emph{2005 IEEE Computer Society Conference on Computer Vision and Pattern Recognition (CVPR'05)}}, Vol.~\bibinfo{volume}{2}. \bibinfo{publisher}{IEEE}, \bibinfo{address}{San Diego, CA, USA}, \bibinfo{pages}{590--596 vol. 2}.
\newblock
\urldef\tempurl%
\url{https://doi.org/10.1109/CVPR.2005.74}
\showDOI{\tempurl}


\bibitem[Gheran et~al\mbox{.}(2018)]%
        {gheran2018gestures}
\bibfield{author}{\bibinfo{person}{Bogdan-Florin Gheran}, \bibinfo{person}{Jean Vanderdonckt}, {and} \bibinfo{person}{Radu-Daniel Vatavu}.} \bibinfo{year}{2018}\natexlab{}.
\newblock \showarticletitle{Gestures for Smart Rings: Empirical Results, Insights, and Design Implications}. In \bibinfo{booktitle}{\emph{Proceedings of the 2018 Designing Interactive Systems Conference}} (Hong Kong, China) \emph{(\bibinfo{series}{DIS '18})}. \bibinfo{publisher}{Association for Computing Machinery}, \bibinfo{address}{New York, NY, USA}, \bibinfo{pages}{623–635}.
\newblock
\showISBNx{9781450351980}
\urldef\tempurl%
\url{https://doi.org/10.1145/3196709.3196741}
\showDOI{\tempurl}


\bibitem[He et~al\mbox{.}(2016)]%
        {he_2016_deep_residual_learning}
\bibfield{author}{\bibinfo{person}{Kaiming He}, \bibinfo{person}{Xiangyu Zhang}, \bibinfo{person}{Shaoqing Ren}, {and} \bibinfo{person}{Jian Sun}.} \bibinfo{year}{2016}\natexlab{}.
\newblock \showarticletitle{Deep Residual Learning for Image Recognition}. In \bibinfo{booktitle}{\emph{Proceedings of the IEEE Conference on Computer Vision and Pattern Recognition (CVPR)}}. \bibinfo{publisher}{IEEE}, \bibinfo{address}{Redmond, WA, USA}, \bibinfo{pages}{770--778}.
\newblock


\bibitem[Iyer et~al\mbox{.}(2020)]%
        {RoboticsPaper}
\bibfield{author}{\bibinfo{person}{Vikram Iyer}, \bibinfo{person}{Ali Najafi}, \bibinfo{person}{Johannes James}, \bibinfo{person}{Sawyer Fuller}, {and} \bibinfo{person}{Shyamnath Gollakota}.} \bibinfo{year}{2020}\natexlab{}.
\newblock \showarticletitle{Wireless steerable vision for live insects and insect-scale robots}.
\newblock \bibinfo{journal}{\emph{Science robotics}} \bibinfo{volume}{5}, \bibinfo{number}{44} (\bibinfo{year}{2020}), \bibinfo{pages}{eabb0839}.
\newblock
\urldef\tempurl%
\url{https://doi.org/10.1126/scirobotics.abb0839}
\showDOI{\tempurl}


\bibitem[Jain et~al\mbox{.}(2023)]%
        {JAIN2023102662}
\bibfield{author}{\bibinfo{person}{Shilpi Jain}, \bibinfo{person}{Sriparna Basu}, \bibinfo{person}{Arghya Ray}, {and} \bibinfo{person}{Ronnie Das}.} \bibinfo{year}{2023}\natexlab{}.
\newblock \showarticletitle{Impact of irritation and negative emotions on the performance of voice assistants: Netting dissatisfied customers’ perspectives}.
\newblock \bibinfo{journal}{\emph{International Journal of Information Management}}  \bibinfo{volume}{72} (\bibinfo{year}{2023}), \bibinfo{pages}{102662}.
\newblock
\showISSN{0268-4012}
\urldef\tempurl%
\url{https://doi.org/10.1016/j.ijinfomgt.2023.102662}
\showDOI{\tempurl}


\bibitem[Ji et~al\mbox{.}(2023)]%
        {ImprovedAlgorithmSmallObjectDetection}
\bibfield{author}{\bibinfo{person}{Shu-Jun Ji}, \bibinfo{person}{Qing-Hua Ling}, {and} \bibinfo{person}{Fei Han}.} \bibinfo{year}{2023}\natexlab{}.
\newblock \showarticletitle{An Improved Algorithm for Small Object Detection Based on YOLO v4 and Multi-scale Contextual Information}.
\newblock \bibinfo{journal}{\emph{Computers and Electrical Engineering}}  \bibinfo{volume}{105} (\bibinfo{year}{2023}), \bibinfo{pages}{108490}.
\newblock
\urldef\tempurl%
\url{https://doi.org/10.1016/j.compeleceng.2022.108490}
\showDOI{\tempurl}


\bibitem[Jing et~al\mbox{.}(2013)]%
        {MagicRingSelfcontainedJingEtAl2013}
\bibfield{author}{\bibinfo{person}{Lei Jing}, \bibinfo{person}{Zixue Cheng}, \bibinfo{person}{Yinghui Zhou}, \bibinfo{person}{Junbo Wang}, {and} \bibinfo{person}{Tongjun Huang}.} \bibinfo{year}{2013}\natexlab{}.
\newblock \showarticletitle{Magic Ring: a self-contained gesture input device on finger}. In \bibinfo{booktitle}{\emph{Proceedings of the 12th International Conference on Mobile and Ubiquitous Multimedia}} (Lule\r{a}, Sweden) \emph{(\bibinfo{series}{MUM '13})}. \bibinfo{publisher}{Association for Computing Machinery}, \bibinfo{address}{New York, NY, USA}, Article \bibinfo{articleno}{39}, \bibinfo{numpages}{4}~pages.
\newblock
\showISBNx{9781450326483}
\urldef\tempurl%
\url{https://doi.org/10.1145/2541831.2541875}
\showDOI{\tempurl}


\bibitem[Kang et~al\mbox{.}(2019)]%
        {MinuetMultimodalInteractionKangEtAl2019a}
\bibfield{author}{\bibinfo{person}{Runchang Kang}, \bibinfo{person}{Anhong Guo}, \bibinfo{person}{Gierad Laput}, \bibinfo{person}{Yang Li}, {and} \bibinfo{person}{Xiang~'Anthony' Chen}.} \bibinfo{year}{2019}\natexlab{}.
\newblock \showarticletitle{Minuet: Multimodal Interaction with an Internet of Things}. In \bibinfo{booktitle}{\emph{Symposium on Spatial User Interaction}} (New Orleans, LA, USA) \emph{(\bibinfo{series}{SUI '19})}. \bibinfo{publisher}{Association for Computing Machinery}, \bibinfo{address}{New York, NY, USA}, Article \bibinfo{articleno}{2}, \bibinfo{numpages}{10}~pages.
\newblock
\showISBNx{9781450369756}
\urldef\tempurl%
\url{https://doi.org/10.1145/3357251.3357581}
\showDOI{\tempurl}


\bibitem[Li et~al\mbox{.}(2012)]%
        {Li2012}
\bibfield{author}{\bibinfo{person}{Xiaoyu Li}, \bibinfo{person}{Shuqin Zeng}, \bibinfo{person}{Yanwei Zhang}, \bibinfo{person}{Ping Wan}, {and} \bibinfo{person}{Jun Wang}.} \bibinfo{year}{2012}\natexlab{}.
\newblock \showarticletitle{Analysis and processing of pixel binning for color image sensor}.
\newblock \bibinfo{journal}{\emph{EURASIP Journal on Advances in Signal Processing}} \bibinfo{volume}{2012}, \bibinfo{number}{1} (\bibinfo{year}{2012}), \bibinfo{pages}{81}.
\newblock
\urldef\tempurl%
\url{https://doi.org/10.1186/1687-6180-2012-81}
\showDOI{\tempurl}


\bibitem[Mayton et~al\mbox{.}(2013)]%
        {WristQuePersonalSensorMaytonEtAl2013}
\bibfield{author}{\bibinfo{person}{Brian~D. Mayton}, \bibinfo{person}{Nan Zhao}, \bibinfo{person}{Matt Aldrich}, \bibinfo{person}{Nicholas Gillian}, {and} \bibinfo{person}{Joseph~A. Paradiso}.} \bibinfo{year}{2013}\natexlab{}.
\newblock \showarticletitle{WristQue: A personal sensor wristband}. In \bibinfo{booktitle}{\emph{2013 IEEE International Conference on Body Sensor Networks}} (Cambridge, MA, USA). \bibinfo{publisher}{IEEE}, \bibinfo{address}{Cambridge, MA, USA}, \bibinfo{pages}{1--6}.
\newblock
\urldef\tempurl%
\url{https://doi.org/10.1109/BSN.2013.6575483}
\showDOI{\tempurl}


\bibitem[McGregor(2024)]%
        {forbes_samsung_galaxy_ring}
\bibfield{author}{\bibinfo{person}{Jay McGregor}.} \bibinfo{year}{2024}\natexlab{}.
\newblock \bibinfo{booktitle}{\emph{Samsung Galaxy Ring: Release Date, Price, Design, Features}}.
\newblock Forbes.
\newblock
\urldef\tempurl%
\url{https://www.forbes.com/sites/jaymcgregor/2024/03/11/samsung-galaxy-ring-release-date-price-design-features/?sh=308cc8f513bf}
\showURL{%
\tempurl}


\bibitem[Miyaoku et~al\mbox{.}(2007)]%
        {CBandFlexibleRingMiyaokuEtAl2007}
\bibfield{author}{\bibinfo{person}{Kento Miyaoku}, \bibinfo{person}{Anthony Tang}, {and} \bibinfo{person}{Sidney Fels}.} \bibinfo{year}{2007}\natexlab{}.
\newblock \showarticletitle{C-Band: A Flexible Ring Tag System for Camera-Based User Interface}. In \bibinfo{booktitle}{\emph{Virtual Reality}}, \bibfield{editor}{\bibinfo{person}{Randall Shumaker}} (Ed.). \bibinfo{publisher}{Springer Berlin Heidelberg}, \bibinfo{address}{Berlin, Heidelberg}, \bibinfo{pages}{320--328}.
\newblock
\showISBNx{978-3-540-73335-5}


\bibitem[Nanayakkara et~al\mbox{.}(2013)]%
        {EyeRingEyeFingerNanayakkaraEtAl}
\bibfield{author}{\bibinfo{person}{Suranga Nanayakkara}, \bibinfo{person}{Roy Shilkrot}, \bibinfo{person}{Kian~Peen Yeo}, {and} \bibinfo{person}{Pattie Maes}.} \bibinfo{year}{2013}\natexlab{}.
\newblock \showarticletitle{EyeRing: a finger-worn input device for seamless interactions with our surroundings}. In \bibinfo{booktitle}{\emph{Proceedings of the 4th Augmented Human International Conference}} (Stuttgart, Germany) \emph{(\bibinfo{series}{AH '13})}. \bibinfo{publisher}{Association for Computing Machinery}, \bibinfo{address}{New York, NY, USA}, \bibinfo{pages}{13–20}.
\newblock
\showISBNx{9781450319041}
\urldef\tempurl%
\url{https://doi.org/10.1145/2459236.2459240}
\showDOI{\tempurl}


\bibitem[Newman(2022)]%
        {fastcompany2022}
\bibfield{author}{\bibinfo{person}{Jared Newman}.} \bibinfo{year}{2022}\natexlab{}.
\newblock \bibinfo{title}{The Smart Home Is Flailing as a Concept Because It Sucks}.
\newblock
\newblock
\urldef\tempurl%
\url{https://www.fastcompany.com/90660570/the-smart-home-is-flailing-as-a-concept-because-it-sucks}
\showURL{%
\tempurl}


\bibitem[Nielsen(1993)]%
        {NielsenResponseTimes}
\bibfield{author}{\bibinfo{person}{Jakob Nielsen}.} \bibinfo{year}{1993}\natexlab{}.
\newblock \bibinfo{title}{{Smart Home Statistics}}.
\newblock \bibinfo{howpublished}{\url{https://www.nngroup.com/articles/response-times-3-important-limits/}}.
\newblock


\bibitem[Oberlo(2024)]%
        {oberlo_2024}
\bibfield{author}{\bibinfo{person}{Oberlo}.} \bibinfo{year}{2024}\natexlab{}.
\newblock \bibinfo{title}{{Smart Home Statistics}}.
\newblock \bibinfo{howpublished}{\url{https://www.oberlo.com/statistics/smart-home-market}}.
\newblock


\bibitem[Oquab et~al\mbox{.}(2024)]%
        {oquab2024dinov2}
\bibfield{author}{\bibinfo{person}{Maxime Oquab}, \bibinfo{person}{Timothée Darcet}, \bibinfo{person}{Théo Moutakanni}, \bibinfo{person}{Huy Vo}, \bibinfo{person}{Marc Szafraniec}, \bibinfo{person}{Vasil Khalidov}, \bibinfo{person}{Pierre Fernandez}, \bibinfo{person}{Daniel Haziza}, \bibinfo{person}{Francisco Massa}, \bibinfo{person}{Alaaeldin El-Nouby}, \bibinfo{person}{Mahmoud Assran}, \bibinfo{person}{Nicolas Ballas}, \bibinfo{person}{Wojciech Galuba}, \bibinfo{person}{Russell Howes}, \bibinfo{person}{Po-Yao Huang}, \bibinfo{person}{Shang-Wen Li}, \bibinfo{person}{Ishan Misra}, \bibinfo{person}{Michael Rabbat}, \bibinfo{person}{Vasu Sharma}, \bibinfo{person}{Gabriel Synnaeve}, \bibinfo{person}{Hu Xu}, \bibinfo{person}{Hervé Jegou}, \bibinfo{person}{Julien Mairal}, \bibinfo{person}{Patrick Labatut}, \bibinfo{person}{Armand Joulin}, {and} \bibinfo{person}{Piotr Bojanowski}.} \bibinfo{year}{2024}\natexlab{}.
\newblock \bibinfo{title}{DINOv2: Learning Robust Visual Features without Supervision}.
\newblock
\newblock
\showeprint[arxiv]{2304.07193}~[cs.CV]


\bibitem[Oura(2024)]%
        {ouraring}
\bibfield{author}{\bibinfo{person}{Oura}.} \bibinfo{year}{2024}\natexlab{}.
\newblock \bibinfo{title}{{Oura Ring}}.
\newblock \bibinfo{howpublished}{\url{https://ouraring.com/}}.
\newblock
\newblock
\shownote{Accessed: March 31, 2024}.


\bibitem[Radford et~al\mbox{.}(2021)]%
        {radford2021learning}
\bibfield{author}{\bibinfo{person}{Alec Radford}, \bibinfo{person}{Jong~Wook Kim}, \bibinfo{person}{Chris Hallacy}, \bibinfo{person}{Aditya Ramesh}, \bibinfo{person}{Gabriel Goh}, \bibinfo{person}{Sandhini Agarwal}, \bibinfo{person}{Girish Sastry}, \bibinfo{person}{Amanda Askell}, \bibinfo{person}{Pamela Mishkin}, \bibinfo{person}{Jack Clark}, \bibinfo{person}{Gretchen Krueger}, {and} \bibinfo{person}{Ilya Sutskever}.} \bibinfo{year}{2021}\natexlab{}.
\newblock \bibinfo{title}{Learning Transferable Visual Models From Natural Language Supervision}.
\newblock
\newblock
\showeprint[arxiv]{2103.00020}~[cs.CV]


\bibitem[Redmon et~al\mbox{.}(2016)]%
        {redmon_2016}
\bibfield{author}{\bibinfo{person}{Joseph Redmon}, \bibinfo{person}{Santosh Divvala}, \bibinfo{person}{Ross Girshick}, {and} \bibinfo{person}{Ali Farhadi}.} \bibinfo{year}{2016}\natexlab{}.
\newblock \showarticletitle{You Only Look Once: Unified, Real-Time Object Detection}. In \bibinfo{booktitle}{\emph{Proceedings of the IEEE Conference on Computer Vision and Pattern Recognition (CVPR)}}. \bibinfo{publisher}{IEEE}, \bibinfo{address}{Seattle, WA, USA}, \bibinfo{pages}{779--788}.
\newblock
\urldef\tempurl%
\url{https://doi.org/10.1109/CVPR.2016.91}
\showDOI{\tempurl}


\bibitem[Reicherts et~al\mbox{.}(2022)]%
        {10.1145/3484221}
\bibfield{author}{\bibinfo{person}{Leon Reicherts}, \bibinfo{person}{Yvonne Rogers}, \bibinfo{person}{Licia Capra}, \bibinfo{person}{Ethan Wood}, \bibinfo{person}{Tu~Dinh Duong}, {and} \bibinfo{person}{Neil Sebire}.} \bibinfo{year}{2022}\natexlab{}.
\newblock \showarticletitle{It's Good to Talk: A Comparison of Using Voice Versus Screen-Based Interactions for Agent-Assisted Tasks}.
\newblock \bibinfo{journal}{\emph{ACM Trans. Comput.-Hum. Interact.}} \bibinfo{volume}{29}, \bibinfo{number}{3}, Article \bibinfo{articleno}{25} (\bibinfo{date}{jan} \bibinfo{year}{2022}), \bibinfo{numpages}{41}~pages.
\newblock
\showISSN{1073-0516}
\urldef\tempurl%
\url{https://doi.org/10.1145/3484221}
\showDOI{\tempurl}


\bibitem[Rodrigues et~al\mbox{.}(2019)]%
        {10.1145/3335595.3335635}
\bibfield{author}{\bibinfo{person}{Ana Rodrigues}, \bibinfo{person}{Rita Santos}, \bibinfo{person}{Jorge Abreu}, \bibinfo{person}{Pedro Be\c{c}a}, \bibinfo{person}{Pedro Almeida}, {and} \bibinfo{person}{S\'{\i}lvia Fernandes}.} \bibinfo{year}{2019}\natexlab{}.
\newblock \showarticletitle{Analyzing the performance of ASR systems: The effects of noise, distance to the device, age and gender}. In \bibinfo{booktitle}{\emph{Proceedings of the XX International Conference on Human Computer Interaction}} (Donostia, Gipuzkoa, Spain) \emph{(\bibinfo{series}{Interacci\'{o}n '19})}. \bibinfo{publisher}{Association for Computing Machinery}, \bibinfo{address}{New York, NY, USA}, Article \bibinfo{articleno}{8}, \bibinfo{numpages}{8}~pages.
\newblock
\showISBNx{9781450371766}
\urldef\tempurl%
\url{https://doi.org/10.1145/3335595.3335635}
\showDOI{\tempurl}


\bibitem[Sapienza(2022)]%
        {brillianttech2022}
\bibfield{author}{\bibinfo{person}{Mia Sapienza}.} \bibinfo{year}{2022}\natexlab{}.
\newblock \bibinfo{title}{Are You Still Relying on Your Phone to Control Your Home?}
\newblock
\newblock
\urldef\tempurl%
\url{https://www.brilliant.tech/blogs/news/are-you-still-relying-on-your-phone-to-control-your-home}
\showURL{%
\tempurl}


\bibitem[Semiconductor(2022)]%
        {nordicsemi_bluetooth_range}
\bibfield{author}{\bibinfo{person}{Nordic Semiconductor}.} \bibinfo{year}{2022}\natexlab{}.
\newblock \bibinfo{booktitle}{\emph{Things You Should Know About Bluetooth Range}}.
\newblock Nordic Semiconductor.
\newblock
\urldef\tempurl%
\url{https://blog.nordicsemi.com/getconnected/things-you-should-know-about-bluetooth-range}
\showURL{%
\tempurl}


\bibitem[Shilkrot et~al\mbox{.}(2015)]%
        {10.1145/2702123.2702421}
\bibfield{author}{\bibinfo{person}{Roy Shilkrot}, \bibinfo{person}{Jochen Huber}, \bibinfo{person}{Wong Meng~Ee}, \bibinfo{person}{Pattie Maes}, {and} \bibinfo{person}{Suranga~Chandima Nanayakkara}.} \bibinfo{year}{2015}\natexlab{}.
\newblock \showarticletitle{FingerReader: A Wearable Device to Explore Printed Text on the Go}. In \bibinfo{booktitle}{\emph{Proceedings of the 33rd Annual ACM Conference on Human Factors in Computing Systems}} (Seoul, Republic of Korea) \emph{(\bibinfo{series}{CHI '15})}. \bibinfo{publisher}{Association for Computing Machinery}, \bibinfo{address}{New York, NY, USA}, \bibinfo{pages}{2363–2372}.
\newblock
\showISBNx{9781450331456}
\urldef\tempurl%
\url{https://doi.org/10.1145/2702123.2702421}
\showDOI{\tempurl}


\bibitem[Stearns et~al\mbox{.}(2018)]%
        {10.1145/3161416}
\bibfield{author}{\bibinfo{person}{Lee Stearns}, \bibinfo{person}{Uran Oh}, \bibinfo{person}{Leah Findlater}, {and} \bibinfo{person}{Jon~E. Froehlich}.} \bibinfo{year}{2018}\natexlab{}.
\newblock \showarticletitle{TouchCam: Realtime Recognition of Location-Specific On-Body Gestures to Support Users with Visual Impairments}.
\newblock \bibinfo{journal}{\emph{Proc. ACM Interact. Mob. Wearable Ubiquitous Technol.}} \bibinfo{volume}{1}, \bibinfo{number}{4}, Article \bibinfo{articleno}{164} (\bibinfo{date}{jan} \bibinfo{year}{2018}), \bibinfo{numpages}{23}~pages.
\newblock
\urldef\tempurl%
\url{https://doi.org/10.1145/3161416}
\showDOI{\tempurl}


\bibitem[Stetten et~al\mbox{.}(2007)]%
        {4371592}
\bibfield{author}{\bibinfo{person}{George Stetten}, \bibinfo{person}{Roberta Klatzky}, \bibinfo{person}{Brock Nichol}, \bibinfo{person}{John Galeotti}, \bibinfo{person}{Kenneth Rockot}, \bibinfo{person}{Kimberly Zawrotny}, \bibinfo{person}{David Weiser}, \bibinfo{person}{Nathan Sendgikoski}, {and} \bibinfo{person}{Samantha Horvath}.} \bibinfo{year}{2007}\natexlab{}.
\newblock \showarticletitle{Fingersight: Fingertip Visual Haptic Sensing and Control}. In \bibinfo{booktitle}{\emph{2007 IEEE International Workshop on Haptic, Audio and Visual Environments and Games}}. \bibinfo{publisher}{IEEE}, \bibinfo{address}{Ottawa, ON, Canada}, \bibinfo{pages}{80--83}.
\newblock
\urldef\tempurl%
\url{https://doi.org/10.1109/HAVE.2007.4371592}
\showDOI{\tempurl}


\bibitem[Store(2024)]%
        {HowSetSmart}
\bibfield{author}{\bibinfo{person}{Google Store}.} \bibinfo{year}{2024}\natexlab{}.
\newblock \bibinfo{booktitle}{\emph{How to Set Up a Smart Home}}.
\newblock Google.
\newblock
\newblock
\shownote{Accessed: March 31, 2024}.


\bibitem[Tan and Le(2019)]%
        {tan_2019_efficientnet}
\bibfield{author}{\bibinfo{person}{Mingxing Tan} {and} \bibinfo{person}{Quoc Le}.} \bibinfo{year}{2019}\natexlab{}.
\newblock \showarticletitle{{E}fficient{N}et: Rethinking Model Scaling for Convolutional Neural Networks}. In \bibinfo{booktitle}{\emph{Proceedings of the 36th International Conference on Machine Learning}} \emph{(\bibinfo{series}{Proceedings of Machine Learning Research}, Vol.~\bibinfo{volume}{97})}, \bibfield{editor}{\bibinfo{person}{Kamalika Chaudhuri} {and} \bibinfo{person}{Ruslan Salakhutdinov}} (Eds.). \bibinfo{publisher}{PMLR}, \bibinfo{address}{Mountain View, CA, USA}, \bibinfo{pages}{6105--6114}.
\newblock
\urldef\tempurl%
\url{https://proceedings.mlr.press/v97/tan19a.html}
\showURL{%
\tempurl}


\bibitem[Through(2022)]%
        {punchthrough_ble_throughput}
\bibfield{author}{\bibinfo{person}{Punch Through}.} \bibinfo{year}{2022}\natexlab{}.
\newblock \bibinfo{booktitle}{\emph{Maximizing BLE Throughput on iOS and Android}}.
\newblock PunchThrough.
\newblock
\urldef\tempurl%
\url{https://punchthrough.com/maximizing-ble-throughput-on-ios-and-android/#:~:text=It%20is%20important%20to%20know,per%20connection%20event%20in%20Android}
\showURL{%
\tempurl}


\bibitem[Ultrahuman(2024)]%
        {ultrahuman}
\bibfield{author}{\bibinfo{person}{Ultrahuman}.} \bibinfo{year}{2024}\natexlab{}.
\newblock \bibinfo{title}{{Ultrahuman}}.
\newblock \bibinfo{howpublished}{\url{https://www.ultrahuman.com//}}.
\newblock
\newblock
\shownote{Accessed: March 31, 2024}.


\bibitem[Ultralytics(2024)]%
        {ultralytics_2024}
\bibfield{author}{\bibinfo{person}{Ultralytics}.} \bibinfo{year}{2024}\natexlab{}.
\newblock \bibinfo{title}{YOLOv8}.
\newblock \bibinfo{howpublished}{\url{https://github.com/ultralytics/yolov8}}.
\newblock
\newblock
\shownote{Accessed: 2024-03-31}.


\bibitem[Vatavu and Bilius(2021)]%
        {vatavu2021gesturing}
\bibfield{author}{\bibinfo{person}{Radu-Daniel Vatavu} {and} \bibinfo{person}{Laura-Bianca Bilius}.} \bibinfo{year}{2021}\natexlab{}.
\newblock \showarticletitle{GestuRING: A Web-based Tool for Designing Gesture Input with Rings, Ring-Like, and Ring-Ready Devices}. In \bibinfo{booktitle}{\emph{The 34th Annual ACM Symposium on User Interface Software and Technology}} (Virtual Event, USA) \emph{(\bibinfo{series}{UIST '21})}. \bibinfo{publisher}{Association for Computing Machinery}, \bibinfo{address}{New York, NY, USA}, \bibinfo{pages}{710–723}.
\newblock
\showISBNx{9781450386357}
\urldef\tempurl%
\url{https://doi.org/10.1145/3472749.3474780}
\showDOI{\tempurl}


\bibitem[Veluri et~al\mbox{.}(2023)]%
        {neuricam}
\bibfield{author}{\bibinfo{person}{Bandhav Veluri}, \bibinfo{person}{Collin Pernu}, \bibinfo{person}{Ali Saffari}, \bibinfo{person}{Joshua Smith}, \bibinfo{person}{Michael Taylor}, {and} \bibinfo{person}{Shyamnath Gollakota}.} \bibinfo{year}{2023}\natexlab{}.
\newblock \bibinfo{booktitle}{\emph{NeuriCam: Key-Frame Video Super-Resolution and Colorization for IoT Cameras}}.
\newblock \bibinfo{publisher}{Association for Computing Machinery}, \bibinfo{address}{New York, NY, USA}, Chapter~25, \bibinfo{pages}{1--17}.
\newblock
\showISBNx{9781450399906}
\urldef\tempurl%
\url{https://doi.org/10.1145/3570361.3592523}
\showURL{%
\tempurl}


\bibitem[Wollner et~al\mbox{.}(2012)]%
        {wollner2012evaluation}
\bibfield{author}{\bibinfo{person}{{P. K.A.} Wollner}, \bibinfo{person}{{P. M.} Langdon}, \bibinfo{person}{T. Goldhaber}, \bibinfo{person}{{I. M.} Hosking}, \bibinfo{person}{A. Mieczakowski}, {and} \bibinfo{person}{{P. J.} Clarkson}.} \bibinfo{year}{2012}\natexlab{}.
\newblock \showarticletitle{Evaluation of setup procedures on mobile devices based on users' initial experience}. In \bibinfo{booktitle}{\emph{NordDesign 2012 - Proceedings of the 9th NordDesign Conference}}, \bibfield{editor}{\bibinfo{person}{{Poul Kyvsgaard} Hansen}, \bibinfo{person}{John Rasmussen}, \bibinfo{person}{{Kaj A.} Jorgensen}, {and} \bibinfo{person}{Christian Tollestrup}} (Eds.). \bibinfo{publisher}{Center for Industrial Production, Aalborg University and Design Society, University of Strathclyde}, \bibinfo{address}{Aalborg, Denmark}, \bibinfo{pages}{1--8}.
\newblock
\newblock
\shownote{9th NordDesign Conference, NordDesign 2012 ; Conference date: 22-08-2012 Through 24-08-2012}.


\bibitem[Yoo et~al\mbox{.}(2015)]%
        {s150714917}
\bibfield{author}{\bibinfo{person}{Yoonjong Yoo}, \bibinfo{person}{Jaehyun Im}, {and} \bibinfo{person}{Joonki Paik}.} \bibinfo{year}{2015}\natexlab{}.
\newblock \showarticletitle{Low-Light Image Enhancement Using Adaptive Digital Pixel Binning}.
\newblock \bibinfo{journal}{\emph{Sensors}} \bibinfo{volume}{15}, \bibinfo{number}{7} (\bibinfo{year}{2015}), \bibinfo{pages}{14917--14931}.
\newblock
\showISSN{1424-8220}
\urldef\tempurl%
\url{https://doi.org/10.3390/s150714917}
\showDOI{\tempurl}


\bibitem[Yoon et~al\mbox{.}(2016)]%
        {TRingInstantCustomizableYoonEtAl2016}
\bibfield{author}{\bibinfo{person}{Sang~Ho Yoon}, \bibinfo{person}{Yunbo Zhang}, \bibinfo{person}{Ke Huo}, {and} \bibinfo{person}{Karthik Ramani}.} \bibinfo{year}{2016}\natexlab{}.
\newblock \showarticletitle{TRing: Instant and Customizable Interactions with Objects Using an Embedded Magnet and a Finger-Worn Device}. In \bibinfo{booktitle}{\emph{Proceedings of the 29th Annual Symposium on User Interface Software and Technology}} (Tokyo, Japan) \emph{(\bibinfo{series}{UIST '16})}. \bibinfo{publisher}{Association for Computing Machinery}, \bibinfo{address}{New York, NY, USA}, \bibinfo{pages}{169–181}.
\newblock
\showISBNx{9781450341899}
\urldef\tempurl%
\url{https://doi.org/10.1145/2984511.2984529}
\showDOI{\tempurl}


\bibitem[Zhang et~al\mbox{.}(2020)]%
        {ThermalRingGestureTagZhangEtAl2020}
\bibfield{author}{\bibinfo{person}{Tengxiang Zhang}, \bibinfo{person}{Xin Zeng}, \bibinfo{person}{Yinshuai Zhang}, \bibinfo{person}{Ke Sun}, \bibinfo{person}{Yuntao Wang}, {and} \bibinfo{person}{Yiqiang Chen}.} \bibinfo{year}{2020}\natexlab{}.
\newblock \showarticletitle{ThermalRing: Gesture and Tag Inputs Enabled by a Thermal Imaging Smart Ring}. In \bibinfo{booktitle}{\emph{Proceedings of the 2020 CHI Conference on Human Factors in Computing Systems}} (Honolulu, HI, USA) \emph{(\bibinfo{series}{CHI '20})}. \bibinfo{publisher}{Association for Computing Machinery}, \bibinfo{address}{New York, NY, USA}, \bibinfo{pages}{1–13}.
\newblock
\showISBNx{9781450367080}
\urldef\tempurl%
\url{https://doi.org/10.1145/3313831.3376323}
\showDOI{\tempurl}


\end{thebibliography}

\end{document}